\documentclass[aps,preprint,prd,nofootinbib]{revtex4-1}
\pdfoutput=1	%arXiv required, within the first 5 lines
\usepackage[utf8]{inputenc}
\usepackage{booktabs}
\usepackage{slashed}
\usepackage{indentfirst}
\usepackage[pdftex]{graphicx}
\usepackage{epstopdf}
\usepackage{amsfonts,amssymb,amsmath}
\usepackage{psfrag}

\usepackage{multirow}
\usepackage{hyperref}
\usepackage{mathrsfs}
\usepackage{url}
\usepackage{appendix}
\usepackage{subfigure}
\usepackage{color}
\usepackage{soul}
\usepackage{ulem}
\usepackage{bm}

\newcommand{\ud}{{\rm{d}}}

\newcommand{\GeV}{{\ \rm{GeV}}}
\newcommand{\MeV}{{\ \rm{MeV}}}

\graphicspath{{./figures/}}

\begin{document}

\title{Rates of  $D^{*}_{0}(2400)$, $ D_J^*(3000) $ as the $D^{*}_{0}(2P)$ and $D^{*}_{0}(3P)$ in $B$ Decays}

\author{Xiao-Ze Tan, Yue Jiang\footnote{jiangure@hit.edu.cn}, Tianhong Wang, Tian Zhou, Geng Li, Zi-Kan Geng, Guo-Li Wang}

\address{Department of Physics, Harbin Institute of
	Technology, Harbin, 150001, People's Republic of China}

\baselineskip=20pt

\vspace*{0.5cm}

\begin{abstract}

In this paper, we use the instantaneous Bethe-Salpeter method to calculate the semi-leptonic and non-leptonic production of the orbitally excited scalar $ D_0^* $ in $ B $ meson decays. When the final state is $ 1P $ state $ D_0^*(2400) $, our theoretical decay rate is consistent with experimental data. For  $ D_J^*(3000) $ final state, which was observed by LHCb collaboration recently and here treated as the orbitally excited scalar $D^{*}_{0}(2P)$, its rate is in the order of $ 10^{-4} \sim 10^{-6}$. We find the special node structure of $D^{*}_{0}(2P)$ wave function possibly results in the suppression of its branching ratio and the abnormal uncertainty. The $3P$ states production rate is in the order of $ 10^{-5}$.

 \vspace*{0.5cm}


\end{abstract}

\maketitle

%%%%%%%%%%%%%%%%%%%%%%%%%%%%%%%%%%%%%%%%%%%%%%%%%%%%%%%%%%%%%%%%%%

%%%%%%%%%%%%%%%%%%%%%%%%%%%%%%%%%%%%%%%%%%%%%%%%%%%%%%%%%%%%%%%%%%

\section{Introduction}

The semi-leptonic and non-leptonic decays of $ B $ mesons are the frequently studied decays and also the dominant production channels of charmed mesons. During the last decades, for many important cases such as providing precise value of CKM element $ V_{cb} $,  the channels of $ B $ decays to S-wave ground states of $ D $ mesons have been extensively measured and studied by the ALEPH, CLEO, OPAL, BABAR and Belle Collaborations\cite{ALEPH1997.373,CLEO1999.3746,OPAL2000.15,Belle2002.258,BABAR2008.032002,BABAR2010.011802,Belle2010.112007,Belle2016.032006} besides the theoretical studies.

In recent years, many collaborations reported several charmed resonances including some orbitally excited $ D $ mesons, which attracts lots of attention.
The Belle and BABAR Collaborations reported the semi-leptonic $ B $ decays to P-wave $ D^* $ mesons by using fully reconstructed $ B $ tags \cite{BABAR2008,Belle2008} and the Belle, BABAR and LHCb Collaborations reported the non-leptonic decays $ B \to D_0^*\pi(K) $ \cite{Belle2004,Belle2007.012006,BABAR2009.112004,LHCb2015.032002,LHCb2015.092002}.
They inspired many theoretical studies on the excited charmed states using different models, for example, the light-front quark model \cite{Kang2018}, the constituent quark model\cite{Segovia2011.094029}, as well as the Bethe-Salpeter method \cite{FuHF2011}, etc.

In 2013, the LHCb Collaboration reported several resonances around 3000 MeV, $ D_J(3000)^0 $ and $ D_J^*(3000)^{+,0} $\cite{LHCb2013}. The $ D_J(3000) $ and $ D_J^*(3000) $ were observed in the $ D^*\pi $ and $D\pi$ invariant mass spectrum respectively. Their quantum numbers $J^P$ are still undetermined and many theoretical studies give different assignments \cite{Sun2013,Lu2014,YuGL2015,Godfrey2016}. In our previous works\cite{lsc2017,lsc2018}, we calculated the strong decays and the leptonic productions of $ D_J(3000)$ and we favoured it as the excited $2P(1^{+'})$ broad state . For $D_J^*(3000)$, we calculated its strong decays and our results favoured it as the excited scalar $2P(0^{+})$ state \cite{Tan2018}.

We notice that, in current experiments and theories, the knowledge of $ B $  semi-leptonic and non-leptonic decays to orbitally excited $ D^{*}_0 $ meson is still rather poor. Thus this work will focus on the leptonic decays $ B \to D_0^* \ell^- \overline{\nu}_{\ell} $ and non-leptonic decays $B \to D_0^* X$, where the initial state could be $B^-$ or $\overline{B}^0$, the final state $D_0^*$ is the excited scalar $D_0^{*}(nP)^0$ or $D_0^{*}(nP)^+$ ( $n=1,~2,~3$), and $X$ is a light meson. Currently, the $1P$ states $D_0^{*}(2400)^0$ and $D_0^{*}(2400)^+$ have been well studied, while $2P$ and $3P$ states haven't. The newly detected $D_J^*(3000)^{+,0}$ are treated as the $2P$ scalars in this paper and our results will help to determine their quantum numbers. The processes of $B$ leptonic decays to them could be their important production ways.

In our previous study \cite{gzk2019}, we found that large relativistic corrections exist in the processes where a heavy-light excited state is involved. We also found that the highly excited state has larger relativistic effect than its corresponding ground state. Thus when a process includes an excited state, a relativistic method or model is needed. In this paper, we use the Bethe-Salper (BS) method based on the relativistic BS equation. The relativistic effect is well concerned by solving the BS equation and applying the BS wave function.

The rest contents of this paper are organized as follows: in section 2, we present the formalism of semi-leptonic production process, including the leptonic and hadronic matrix elements by using the BS method. Then the factorization approach is used to derive the formalism of non-leptonic process in section 3. In section 4, we show our numerical results and comparison with the results of other model. Finally, discussions and short summary are given in section 4.

\section{Formalism of semi-leptonic decays}

\begin{figure}[htb!]
	\centering
	\includegraphics[scale=0.5]{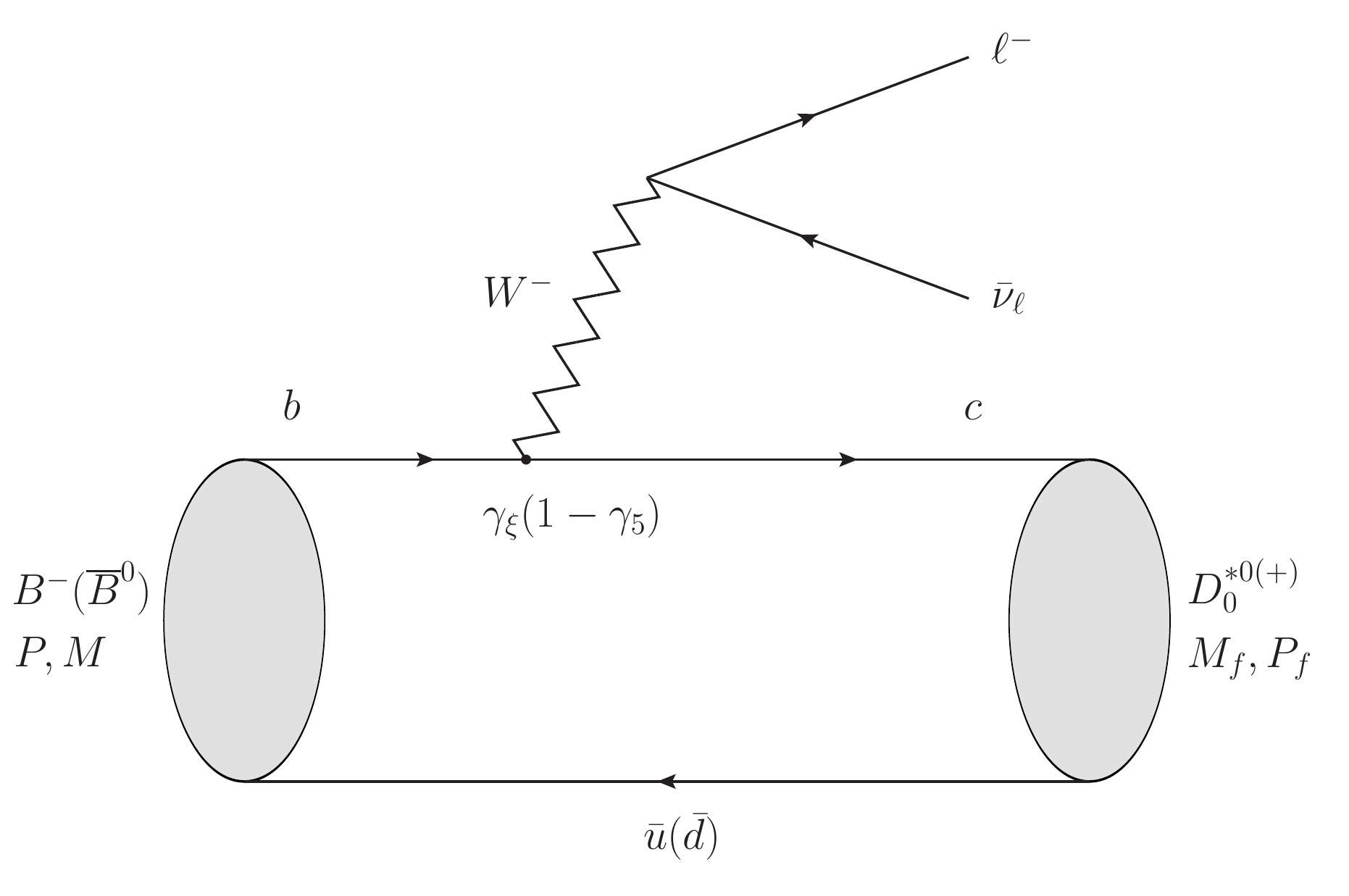}
	\caption[fig1]{Feynman diagram of the semi-leptonic decay $B^{-}(\overline{B}^0)\to D_0^{*0(+)} \ell^- \overline{\nu}_{\ell}$.}
	\label{feynman-sl}
\end{figure}

We take $ B^{-}(\overline{B}^0) \to D_{0}^{*0}(D_{0}^{*+})\ell^- \overline{\nu}_{\ell}$ as an example to show the calculation details of semi-leptonic process. The Feynman diagram is shows in Fig. \ref{feynman-sl}.

The transition amplitude $T  $ can be expressed as :
\begin{equation}
	T=\frac{G_F}{\sqrt{2}}V_{cb} l^{\xi} \left< D_0^*(P_f) | J_{\xi} | B^-(P) \right>,
\end{equation}
where $ G_F $ is the Fermi weak coupling constant, $ V_{cb} $ is the CKM matrix element, $J_{\xi}$ is the charged weak current and $ l^{\xi}  $ is the leptonic matrix.
\begin{align}
	& J_{\xi} = \bar{c} \gamma_{\xi}(1-\gamma_{5})b , \\
	& l^{\xi} = \bar{u}_{\ell}(p_l)\gamma^{\xi}(1-\gamma_5){v_{\nu}}(p_{{\nu}}) .
\end{align}

Then the square of amplitude can be expressed by the function of hadronic and leptonic tensor,
\begin{equation}
	|T|^2 = \frac{G_F^2}{2}|V_{cb}|^2 l^{\xi \xi '} h_{\xi \xi '},
\end{equation}
where the leptonic tensor can be written as following form:
\begin{equation}
\begin{split}
l^{\xi \xi '} = 8(p_{\nu}^{\xi} p_{\ell}^{\xi '} + p_{\ell}^{\xi}p_{\nu}^{\xi '} - p_{\ell} \cdot p_{\nu} g^{\xi \xi '} + i \varepsilon ^{\xi \xi' p_{\ell} p_{\nu}}).
\end{split}
\end{equation}

We derive the hadronic matrix element by using the relativistic BS method
\begin{equation}
	\begin{split}
		h_{\xi}&=\left< D_0^*(P_f) | J_{\xi} | B^-(P) \right> \\
		 &=  \int \frac{\ud^3 q}{(2\pi)^3} \mathrm{Tr}\left[ \frac{\slashed{P}}{M} \overline{\varphi}_{P_f}(q_{f\perp}) \gamma_{\xi} (1-\gamma_{5}) \varphi_{P}(q_{\perp}) \right] = n_1 P_{\xi} + n_2 P_{f\xi} ,
	\end{split}
\end{equation}
where $q_{f}=q-\frac{m_u}{m_b+m_u}P_{f}$, $ \varphi_{P}  $ and $ \varphi_{P_f}  $ are the instantaneous BS wave functions of the initial state $ B $ meson and final state $ D_0^* $ mesons. They are obtained by completely solving the BS equation. The processes of solving the Salpeter equation and obtaining the wave functions are not shown here. More details can be found in our previous works \cite{WangGL2004,WangGL2006, WangGL2007}. We just give a brief review in the appendix.  $ n_i $ are the form factors whose results are shown in next section.

$ h_{\xi \xi '} $ is the hadronic tensor,
\begin{equation}
\label{htesor}
	\begin{split}
	h_{\xi \xi '} =& \left< D_0^*(P_f) | J_{\xi} | B^-(P) \right>^{\dagger} \left< D_0^*(P_f) | J_{\xi '} | B^-(P) \right> \\
	=& n_1^2 P_{\xi}P_{ \xi '} + n_2^2 P_{f \xi }P_{f \xi '} + n_1n_2 (P_{\xi}P_{f\xi '}+P_{f \xi} P_{\xi '}).
	\end{split}
\end{equation}

Then the decay width can be given by the phase-space integral
\begin{equation}
	\Gamma=\frac{1}{2M} \int \frac{\ud^3 \bm{P}_f \ud^3 \bm{p}_{\ell} \ud^3 \bm{p}_{{\nu}_{\ell}}}{(2\pi)^9 2E_f 2E_{\ell} 2E_{{\nu}_{\ell}}} (2\pi)^4\delta^4(P-P_f-p_{\ell}-p_{{\nu}_{\ell}})  |T|^2.
\end{equation}
After the simplification, it can be rewritten as
\begin{equation}
	\Gamma=\frac{1}{64\pi M } \int   |T|^2  \frac{|\bm{P}_f|}{E_f} \ud |\bm{P}_f |  \frac{|\bm{p}_{\ell}|}{E_{\ell}} \ud |\bm{p}_{\ell} |.
\end{equation}

\section{Formalism of non-leptonic decays}

\begin{figure}[htb!]
	\centering
	\includegraphics[scale=0.5]{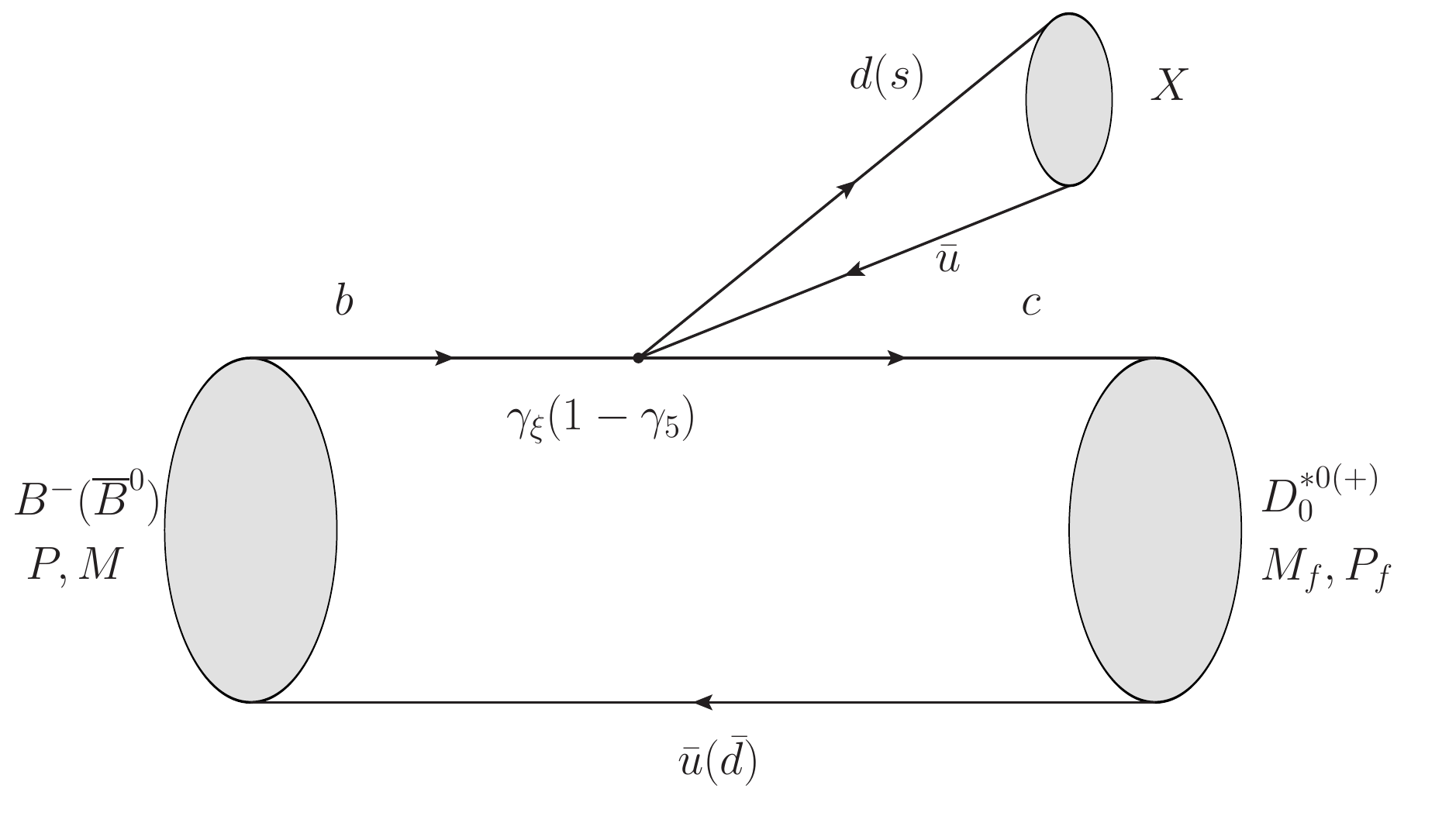}
	\caption[fig2]{Feynman diagram of the non-leptonic decay $B^{-}(\overline{B}^0)\to D_0^{*0(+)} X$.}
	\label{feynman-nl}
\end{figure}

The Feynman diagram of non-leptonic decay $B^{-}\to D_0^{*0} X$ or $\overline{B}^{0}\to D_0^{*+} X$ is shown in Fig. \ref{feynman-nl}, where $ X $ could be $ \pi^-,~K^-, ~\rho^-$ or $ K^{*-} $.

The effective Hamiltonian $\mathcal{H}_{eff}  $ can be expressed as \cite{Ali1998.094009,Choi2009.114003}:
\begin{equation}
	\begin{split}
	\mathcal{H}_{eff} = \frac{G_F}{2}\left[ V_{cb}V^*_{uq}(c_1(\mu) O_1 + c_2(\mu) O_2) + h.c.  \right],
	\end{split}
\end{equation}
where, $ O_i $ is the 4-quark operator containing the charged weak current $ (\bar{q}u)_{V-A} $ and $ (\bar{c}b)_{V-A} $; $ c_i $ is the Wilson coefficient which depends on the renormalization scale $ \mu $.

By using the factorization approach, the transition matrix elements$ \left< D_0^*X|\mathcal{H}_{eff}|B \right> $ involving the 4-quark operators can be split into the product of two matrix elements $\left< D_0^*|(\bar{c}b)_{V-A}|B \right> $ and $ \left< X|(\bar{q}u)_{V-A}|0 \right> $ \cite{Fakirov1978.315,Bauer1987.103,Ali1998.094009}.

Then the transition amplitude can be expressed as
\begin{equation}
	T=\frac{G_F}{\sqrt{2}} V_{cb} V^*_{uq} a_1(\mu) \left<D_0^{*0} | (\bar{c}b)_{V-A} | B^- \right> \left< X|(\bar{q}u)_{V-A}|0 \right>,
\end{equation}
where, $ q $ denotes the $ d $ or $ s $ quark; $ a_1=c_1+\frac{1}{N_c}c_2 $ is the effective Wilson coefficient, where $ N_c=3 $ is the number of colors. We choose the scale $ \mu \approx m_b$ for $ B $ decays and adopt the effective Wilson coefficient $ a_1=1.14 $ \cite{Ivanov2006.054024}. The annihilation matrix element can be written as
\begin{equation}
	\begin{split}
	\left< X_P | (\bar{q}u)^{\mu}_{V-A} |0 \right> &= -i f_{X} P_{X}^{\mu}, \\
	\left< X_V | (\bar{q}u)^{\mu}_{V-A} |0 \right> &=  f_{X} M_{X}\varepsilon_{X}^{\mu},
	\end{split}
\end{equation}
where $ X_P $ means pseudoscalar ($ \pi , K $) and $ X_V $ means vector mesons ($ \rho , K^* $);  $ f_X $ is the corresponding decay constant; $ \varepsilon^{\mu} $ is the polarization vector of $ X_V $ and it satisfies the completeness relation $  \sum\limits \varepsilon_{\lambda}^{\mu} \varepsilon_{\lambda}^{\nu}=\frac{P_X^{\mu}P_X^{\nu}}{M_X^2}-g^{\mu\nu}$.

Same as the semi-leptonic case, we write the square of amplitude by the hadronic and light meson tensor
\begin{equation}
	|T|^2 = \frac{G_F^2}{2}|V_{ub}|^2 |V_{uq}|^2 a_1^2 h_{\mu\nu} X^{\mu\nu},
\end{equation}
where $ h_{\mu\nu} $ is same as Eq. \ref{htesor}, and the light meson tensor is
\begin{equation}
	\begin{split}
	X_P^{\mu\nu} &= P_X^{\mu}P_X^{\nu}f_X^2,	\\
	X_V^{\mu\nu} &= (P_X^{\mu}P_X^{\nu} - M_X^2 g^{\mu\nu}) f_X^2.
	\end{split}
\end{equation}

Then the decay width can be obtained by
\begin{equation}
	\Gamma = \int \frac{|\bm{P}_f|}{32\pi^2 M^2} \sum |T|^2 \ud \Omega.
\end{equation}

\section{RESULTS AND DISCUSSION}

In our calculations, we adopt the same parameters as what we used before\cite{Tan2018}: $m_u=0.305\GeV,\ m_d=0.311\GeV,\ m_s=0.50\GeV,\   m_c=1.62\GeV$,  $\alpha=0.060\GeV$, $\lambda=0.210\GeV^2$ and $\Lambda_{QCD}=0.270 \GeV$. The involved mesons' masses are: $M_{D_0^*(2400)^0}=2.318\GeV $, $M_{D_0^*(2400)^+}=2.351 \GeV$,
 ${ M_{D_J^*(3000)^{(0,+)}}=3.008 \GeV}$, $ M_{D_0^*(3P)^{0,+}}=3.183\GeV$, $M_{B^0}=5.2796\GeV$ and  $ M_{B^{\pm}}=5.2793\GeV$.

The CKM matrix elements\cite{PDG2018} and the involved mesons' decay constants are\cite{PDG2018,LHCb2013,LiQ2017}: $ |V_{cb}|=0.0422 $, $ |V_{ud}|=0.9742 $, $ |V_{us}|=0.2243 $, 
$f_\rho=205\MeV $, $f_\pi=130.4\MeV$, $   f_\eta=130\MeV$ and $  f_k=156.2\MeV$.

\subsection{Semi-leptonic decays}

\begin{figure}[htb!]
	\centering
	\subfigure[The form factors for $D_0^*(2400),\ l=e $ or $\mu $]{
		\label{ff2400e} %% label for first subfigure
		\includegraphics[width=0.45\textwidth]{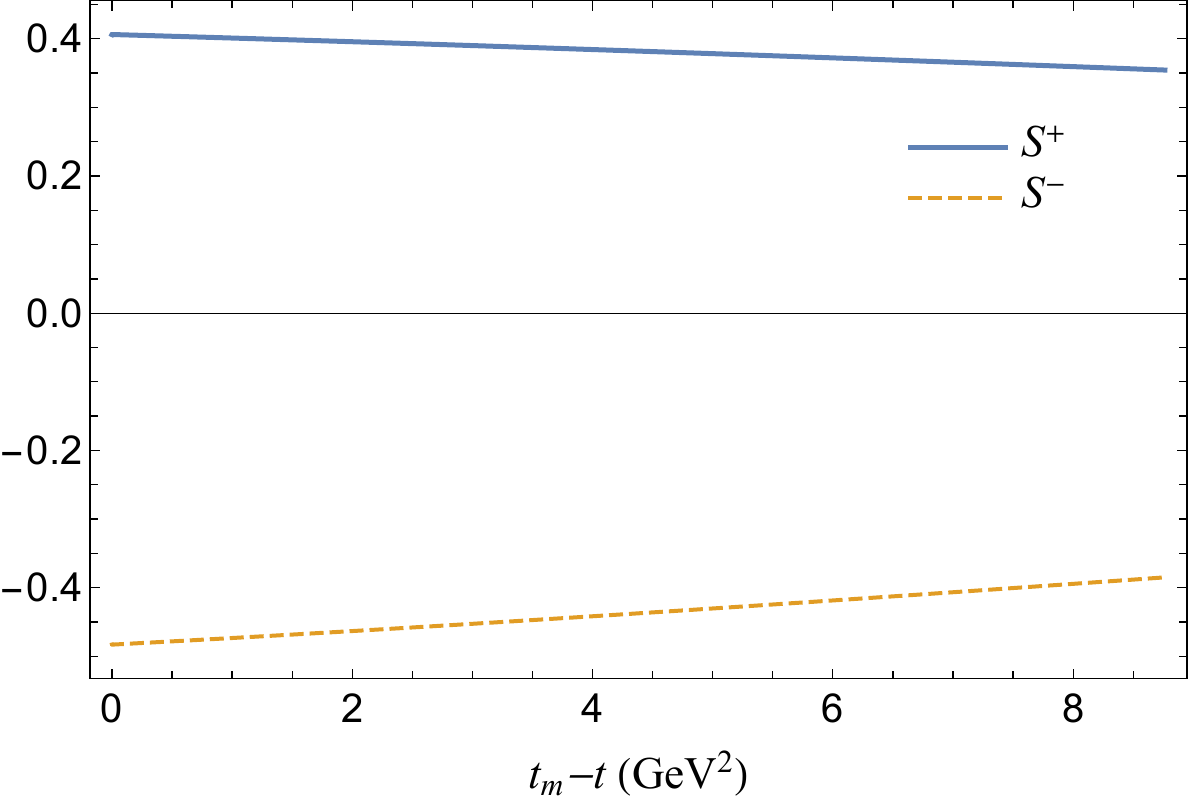}}
	\hspace{0.2cm}
	\subfigure[The form factors for $D_0^*(2400),\ l=\tau $]{
		\label{ff2400tau} %% label for first subfigure
		\includegraphics[width=0.45\textwidth]{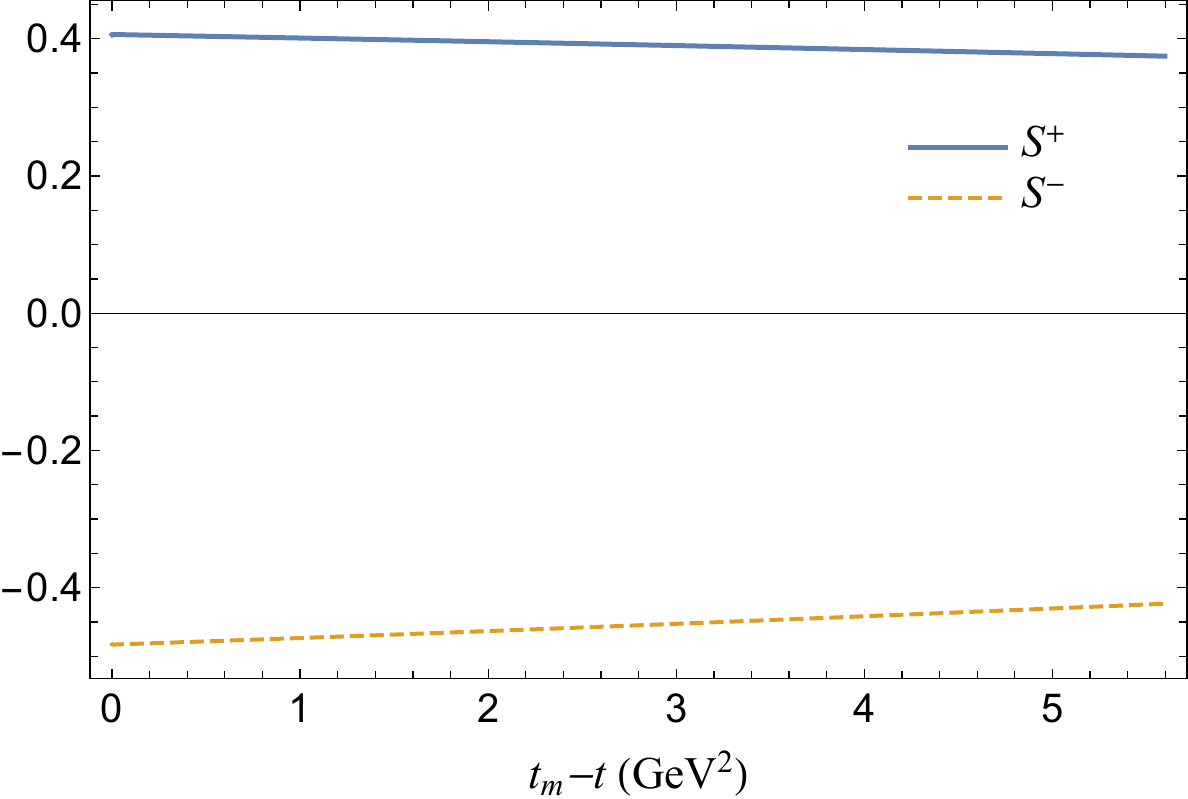}}
	
	\subfigure[The form factors for $D_J^*(3000),\ l=e$ or $\mu $]{
		\label{ff3000e} %% label for first subfigure
		\includegraphics[width=0.45\textwidth]{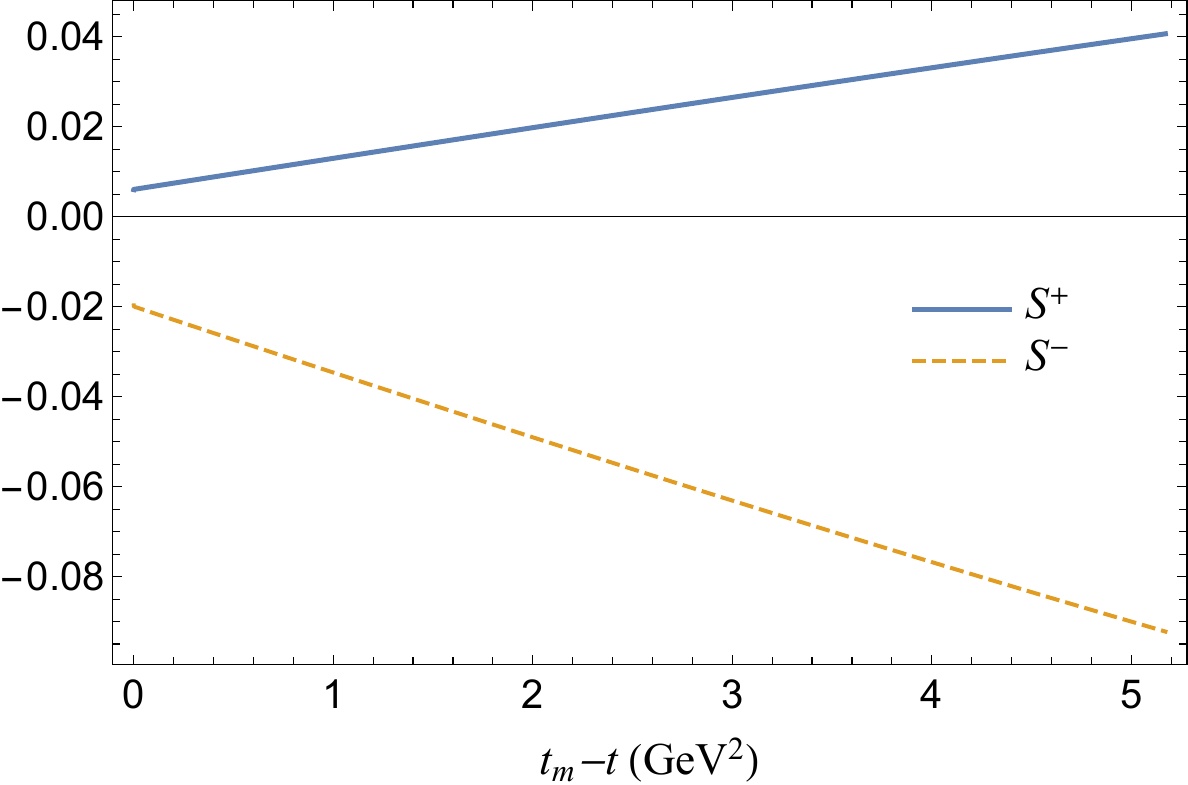}}
	\hspace{0.2cm}
	\subfigure[The form factors for $D_J^*(3000),\ l=\tau $]{
		\label{ff3000tau} %% label for first subfigure
		\includegraphics[width=0.45\textwidth]{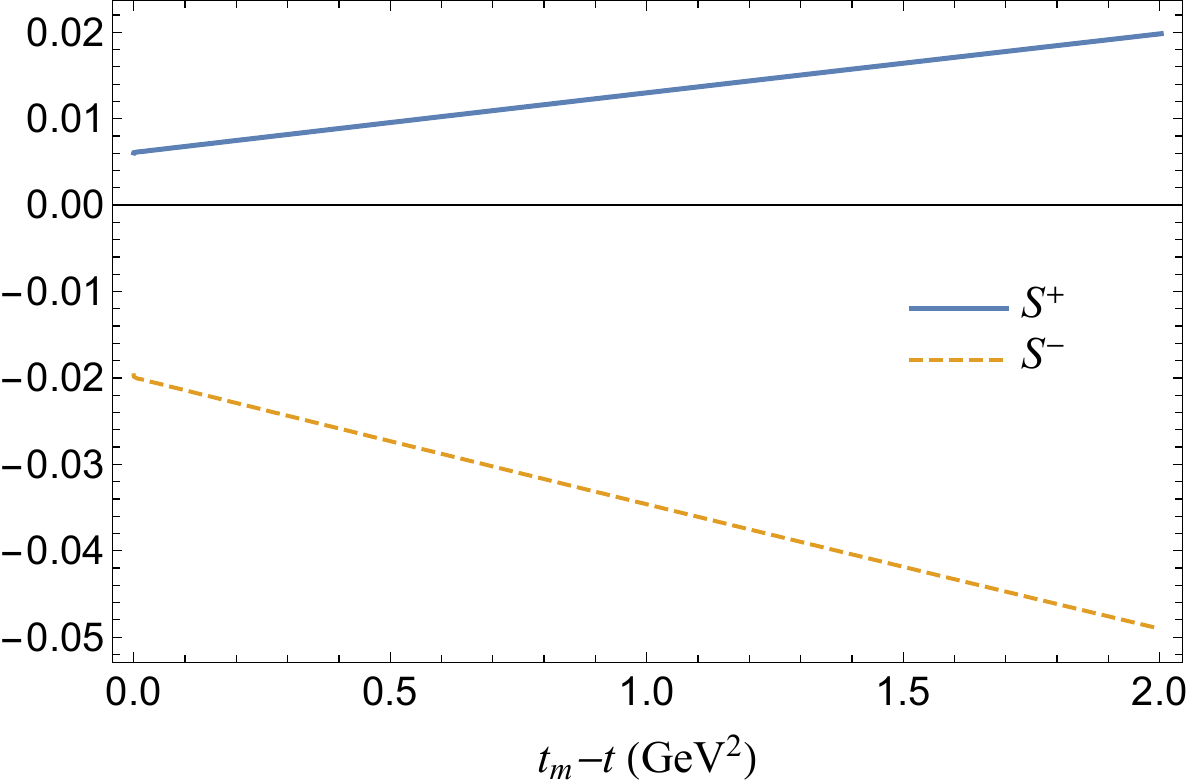}}
	
	\subfigure[The form factors for $D_0^*(3P),\ l=e$ or $\mu $]{
		\label{ff3183e} %% label for first subfigure
		\includegraphics[width=0.45\textwidth]{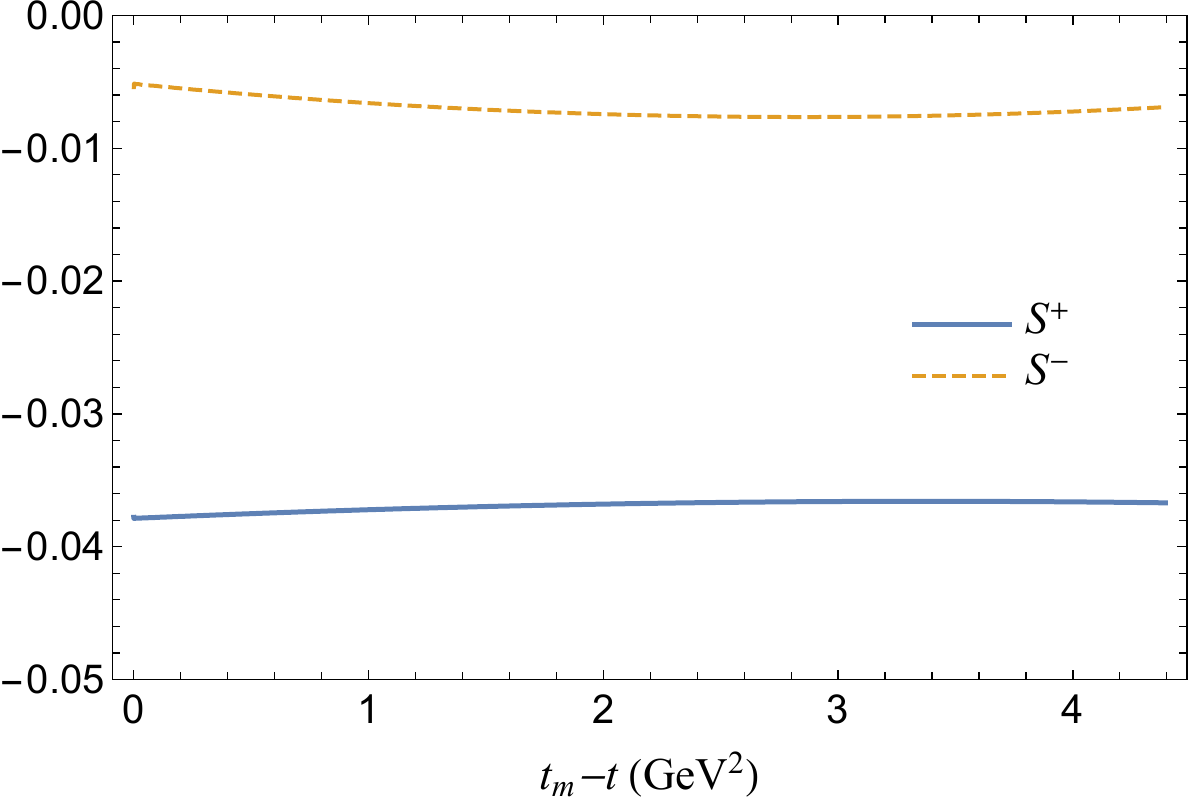}}
	\hspace{0.2cm}
	\subfigure[The form factors for $D_0^*(3P),\ l=\tau $]{
		\label{ff3183tau} %% label for first subfigure
		\includegraphics[width=0.45\textwidth]{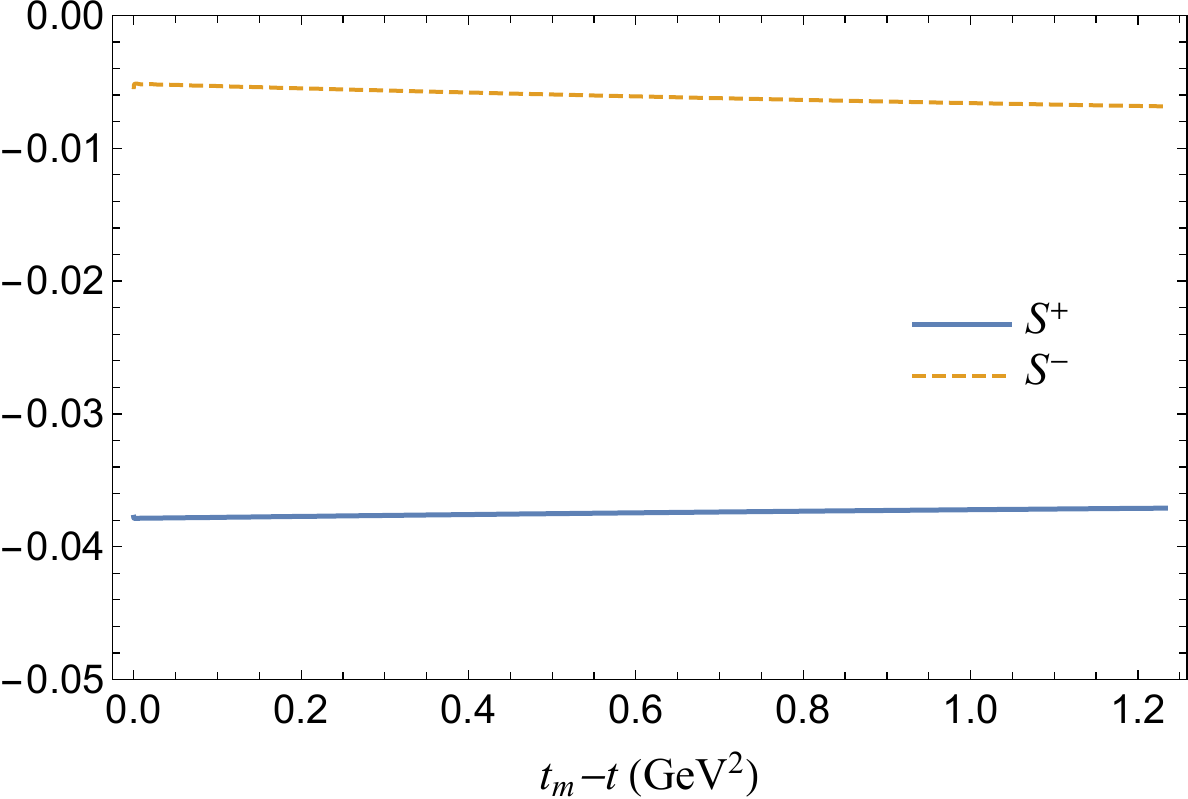}}
	\caption{The form factors for the semi-leptonic production process of $D_0^*(2400)$ , $D_J^*(3000)$ and $ D_0^*(3P) $ , where $ S^+=(n_1+n_2)/2,\ S^-=(n_1-n_2)/2 $.}
	\label{ff}
\end{figure}

The form factors relevant to the hadronic transition matrix elements of $B\to D_0^*(1P-3P) \ell \nu_{\ell}$ are shown in Fig. \ref{ff}, where $ t=(P-P_f)^2 $ and  $ t_m $ is the momentum transfer at the zero recoil(the maximum value of $t$).

Table \ref{slresuluts2400} shows the decay widths and  branching ratios of semi-leptonic production of the ground state $ D_0^*(2400) $.  With varying the parameters by $ \pm 5 \% $, we can obtain the uncertainty of the results. Because there is almost no difference between the results of $ l=e $ and $ l=\mu $, only the values of $ l=e $ are given below.

\begin{table}[htb!]
	\renewcommand\arraystretch{0.99}
	\caption[results]{The results of semi-leptonic production of $ 1P $ state $ D_0^*(2400) $ }
	\label{slresuluts2400}
	\vspace{0.5em}\centering
	\begin{tabular}{lcc}
		\toprule[1.5pt]
		\multicolumn{1}{c}{Channels} & Widths (GeV) & BR  \\
		\midrule[1pt]
		$ B^{-} \to D_0^*(2400)^0 e^- \overline{\nu}_e $ & $1.03 \sim 1.70  \times 10^{-15} $ &  $ 2.57 \sim 4.24 \times 10^{-3} $ \\
		 $ \overline{B}^{0} \to D_0^*(2400)^+ e^- \overline{\nu}_e $ & $1.04 \sim 1.69  \times 10^{-15} $ &  $ 2.39 \sim 3.91 \times 10^{-3} $  \\
		$ B^{-} \to D_0^*(2400)^0 \tau^- \overline{\nu}_{\tau} $ & $ 1.20 \sim 1.72 \times 10^{-16} $ & $ 2.98 \sim 4.28 \times 10^{-4} $ \\
		 $ \overline{B}^{0} \to D_0^*(2400)^+ \tau^- \overline{\nu}_{\tau} $ & $ 1.13 \sim 1.60 \times 10^{-16} $ & $ 2.62 \sim 3.70 \times 10^{-4} $ \\
		\bottomrule[1.5pt]
	\end{tabular}
\end{table}

For comparison, we also give the results of cascade decays to $ D\pi  $ which are shown in Table \ref{cascadesl}.  Recently, Ref.\cite{Kang2018}  used covariant light-front quark model to calculate the channel $ B^{+} \to D_0^*(2400)^0 l^+ \overline{\nu}_l   \to D^- \pi^+ $ , whose result of branching ratio is $ 2.31\pm 0.25 \times 10^{-3} $. And Ref.\cite{Segovia2011.094029} shows that $ \mathcal{B}(B^{+} \rightarrow \overline{D}_{0}^{* 0} l^{+} \nu_{l}) \times \mathcal{B}(\overline{D}_{0}^{* 0} \rightarrow D^{-} \pi^{+})=2.15 \times 10^{-3}  $ and $ \mathcal{B}\left(B^{0} \rightarrow D_{0}^{*-} l^{+} \nu_{l}\right) \times \mathcal{B}\left(D_{0}^{*-} \rightarrow D^{0} \pi^{-}\right)=1.80 \times 10^{-3} $ with the constituent quark model.
Considering the uncertainty, our results are consistent with the experimental and other models' results.

\begin{table}[htb!]
	\renewcommand\arraystretch{0.90}
	\caption[results]{The branching ratios($ \times 10^{-3} $)  of cascade decays to $ D \pi $(with $ l=e $).  The branching ratio $\mathcal{B}( D_0^*(2400) \to D \pi ) $ is from our previous work\cite{Tan2018}. }
	\label{cascadesl}
	\vspace{0.5em}\centering
	\begin{tabular}{lcccc}
		\toprule[1.5pt]
		\multicolumn{1}{c}{Channels} &  Ours  &  BABAR\cite{BABAR2008}   &   Belle\cite{Belle2008}   & PDG\cite{PDG2018}  \\
		\midrule[1pt]
		$ \mathcal{B}  (B^{-} \to D_0^*(2400)^0 e^- \overline{\nu}_l  \to D^0 \pi^0 )$ &  $ 0.85 \sim 1.39  $ & $ $ &  $  $   \\
		$ \mathcal{B}( B^{-} \to D_0^*(2400)^0 e^- \overline{\nu}_l   \to D^+ \pi^- )$ &  $ 1.72 \sim 2.81   $ & $ 2.6 \pm 0.9  $ &  $ 2.4 \pm 1.0  $ & $ 2.5\pm 0.5  $  \\
		$ \mathcal{B}( \overline{B}^{0} \to D_0^*(2400)^+ e^- \overline{\nu}_l   \to D^0 \pi^+ )$ &  $ 1.60 \sim 2.61  $ &  $ 4.4 \pm 1.4  $ &  $ 2.0 \pm 1.2  $  & $ 3.0\pm 1.2 $ \\
		$ \mathcal{B}( \overline{B}^{0} \to D_0^*(2400)^+ e^- \overline{\nu}_l   \to D^+ \pi^0 )$ &  $ 0.79 \sim 1.29  $  & $ $ &  $  $  \\
		\bottomrule[1.5pt]
	\end{tabular}
\end{table}

Then we use the same method to calculate the $2P$ state $D_J(3000)$, which is shown in Table \ref{slresuluts3000}. Unlike the ground state, the results of $ 2P $ state are much lower and have large uncertainty by varying the input parameters, while the predicted results of $ 3P $ state are given in Table \ref{slresuluts3183} and get smaller uncertainty.

\begin{table}[htb!]
	\renewcommand\arraystretch{1.01}
	\caption[results]{The results of semi-leptonic production of  $ 2P $ state $ D_J^*(3000) $ }
	\label{slresuluts3000}
	\vspace{0.5em}\centering
	\begin{tabular}{lcc}
		\toprule[1.5pt]
		\multicolumn{1}{c}{Channels} & Widths (GeV) & BR   \\
		\midrule[1pt]
		$ B^{-} \to D_J^*(3000)^0 e^- \overline{\nu}_e $ & $ 1.6	\sim 80.2 \times 10^{-18} $ & $ 0.4	\sim 20.0 \times 10^{-5} $ \\
		 $ \overline{B}^{0} \to D_J^*(3000)^+ e^- \overline{\nu}_e $ & $ 1.2 \sim 83.0 \times 10^{-18} $ & $ 0.3 \sim 19.2 \times 10^{-5} $  \\
		$ B^{-} \to D_J^*(3000)^0 \tau^- \overline{\nu}_{\tau} $ & $ 0.4 \sim 69.8 \times10^{-20} $ & $ 0.1 \sim 17.3 \times 10^{-7} $ \\
		 $ \overline{B}^{0} \to D_J^*(3000)^+ \tau^- \overline{\nu}_{\tau} $ & $ 0.8 \sim 72.9\times10^{-20} $ & $ 0.2 \sim 16.8 \times 10^{-7} $ \\
		\bottomrule[1.5pt]
	\end{tabular}
\end{table}

\begin{table}[htb!]
	\renewcommand\arraystretch{0.99}
	\caption[results]{The results of semi-leptonic production of $ 3P $ state $ D_0^*$ }
	\label{slresuluts3183}
	\vspace{0.5em}\centering
		\begin{tabular}{lcc}
			\toprule[1.5pt]
			\multicolumn{1}{c}{Channels} & Widths (GeV) & BR   \\
			\midrule[1pt]
			$ B^{-} \to D_0^*(3P)^0 e^- \overline{\nu}_e $ & $3.10 \sim 7.37  \times 10^{-18} $ &  $ 0.77 \sim 1.83\times 10^{-5} $ \\
			 $ \overline{B}^{0} \to D_0^*(3P)^+ e^- \overline{\nu}_e $ & $ 3.24 \sim 7.88  \times 10^{-18} $ &  $ 0.75 \sim 1.82 \times 10^{-5} $  \\
			$ B^{-} \to D_0^*(3P)^0 \tau^- \overline{\nu}_{\tau} $ & $ 2.49 \sim 3.67 \times 10^{-20} $ & $ 6.19 \sim 9.13 \times 10^{-8} $ \\
			 $ \overline{B}^{0} \to D_0^*(3P)^+ \tau^- \overline{\nu}_{\tau} $ & $ 2.45 \sim 3.78 \times 10^{-20} $ & $ 5.65 \sim 8.73 \times 10^{-8} $ \\
			\bottomrule[1.5pt]
	\end{tabular}
\end{table}

\begin{figure}[htb!]
	\centering
	\subfigure[Wave functions of  $ B^- $ ]{
		\label{0-wf} %% label for first subfigure
		\includegraphics[width=0.45\textwidth]{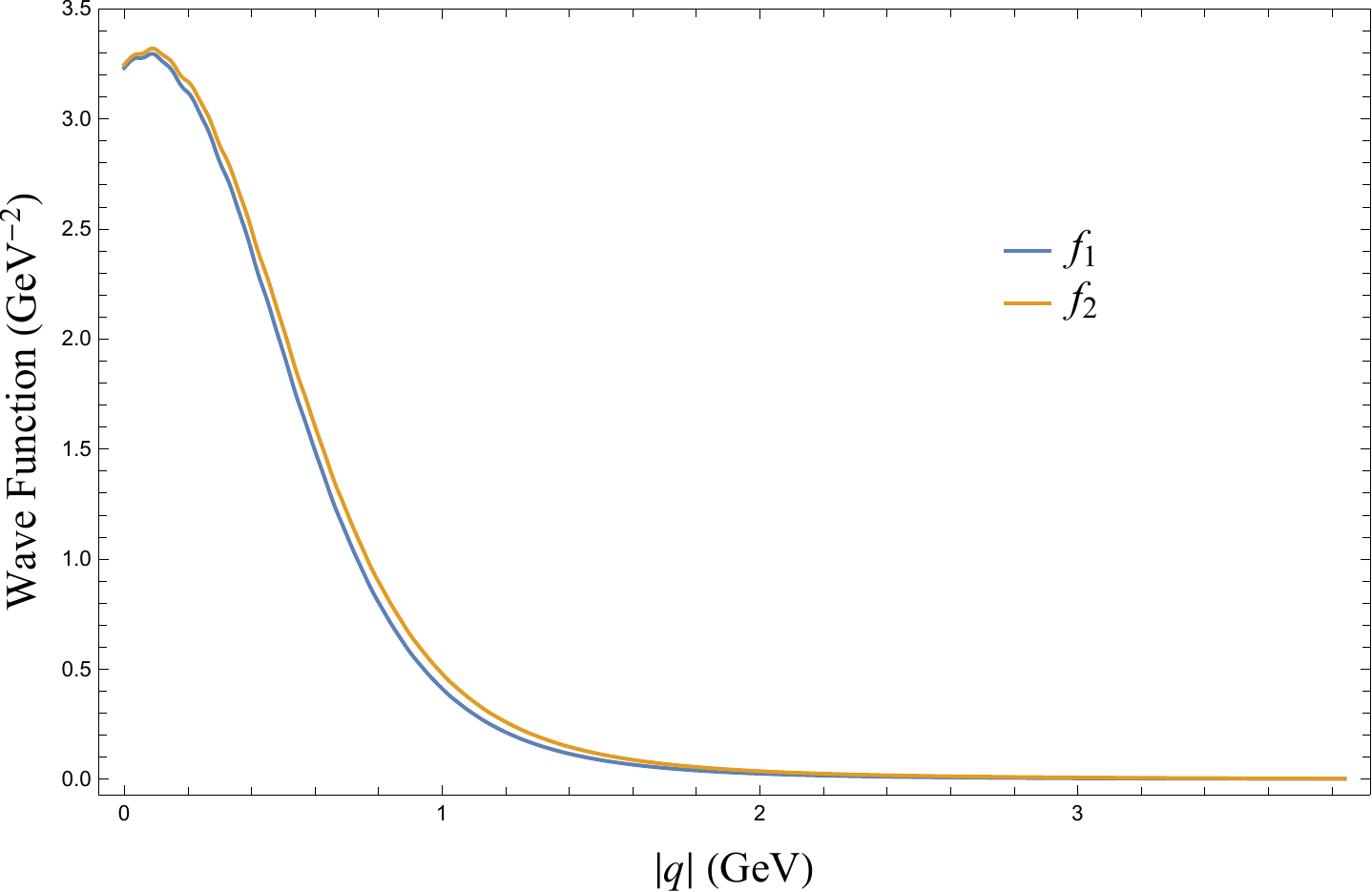}}
	\hspace{0.2cm}
	\subfigure[Wave functions of $ 1P $ state $ D_0^*(2400) $]{
		\label{1Pwf} %% label for first subfigure
		\includegraphics[width=0.45\textwidth]{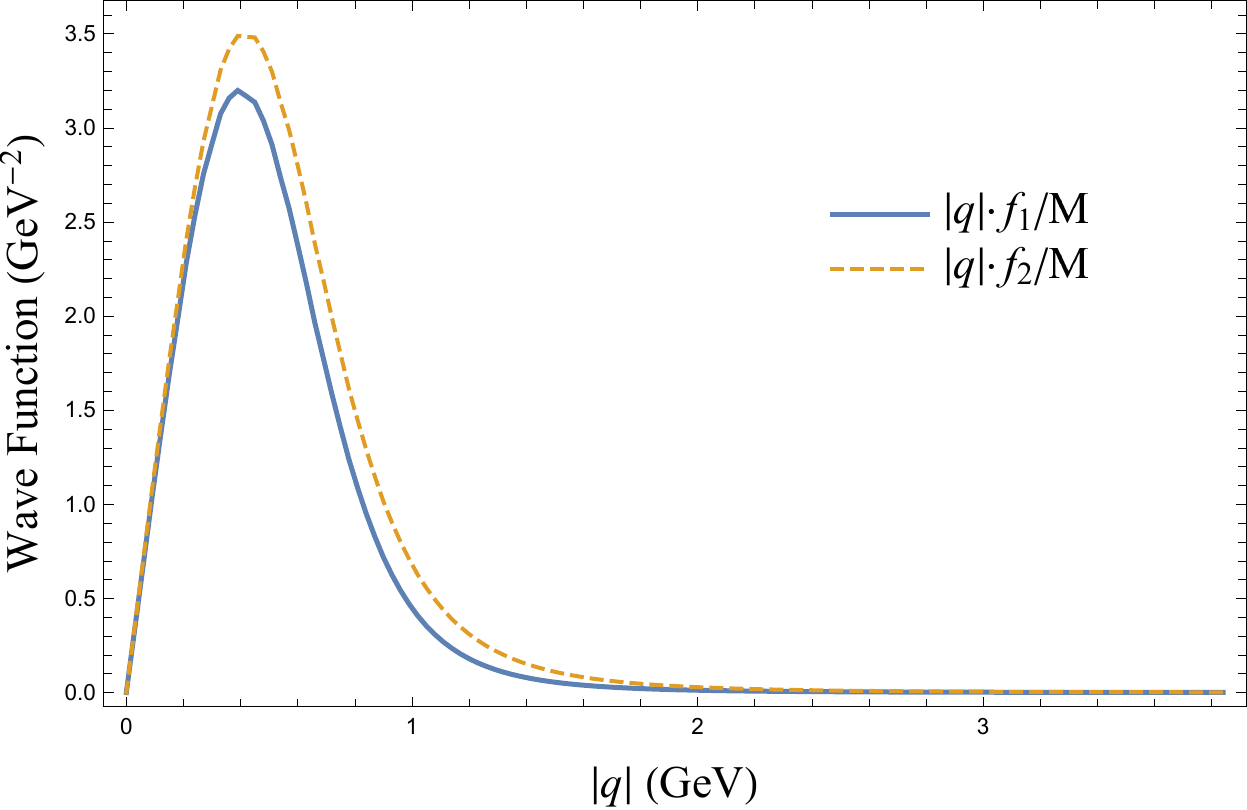}}
	
	\subfigure[Wave functions of $ 2P $ state $ D_J^*(3000) $]{
		\label{2Pwf} %% label for first subfigure
		\includegraphics[width=0.45\textwidth]{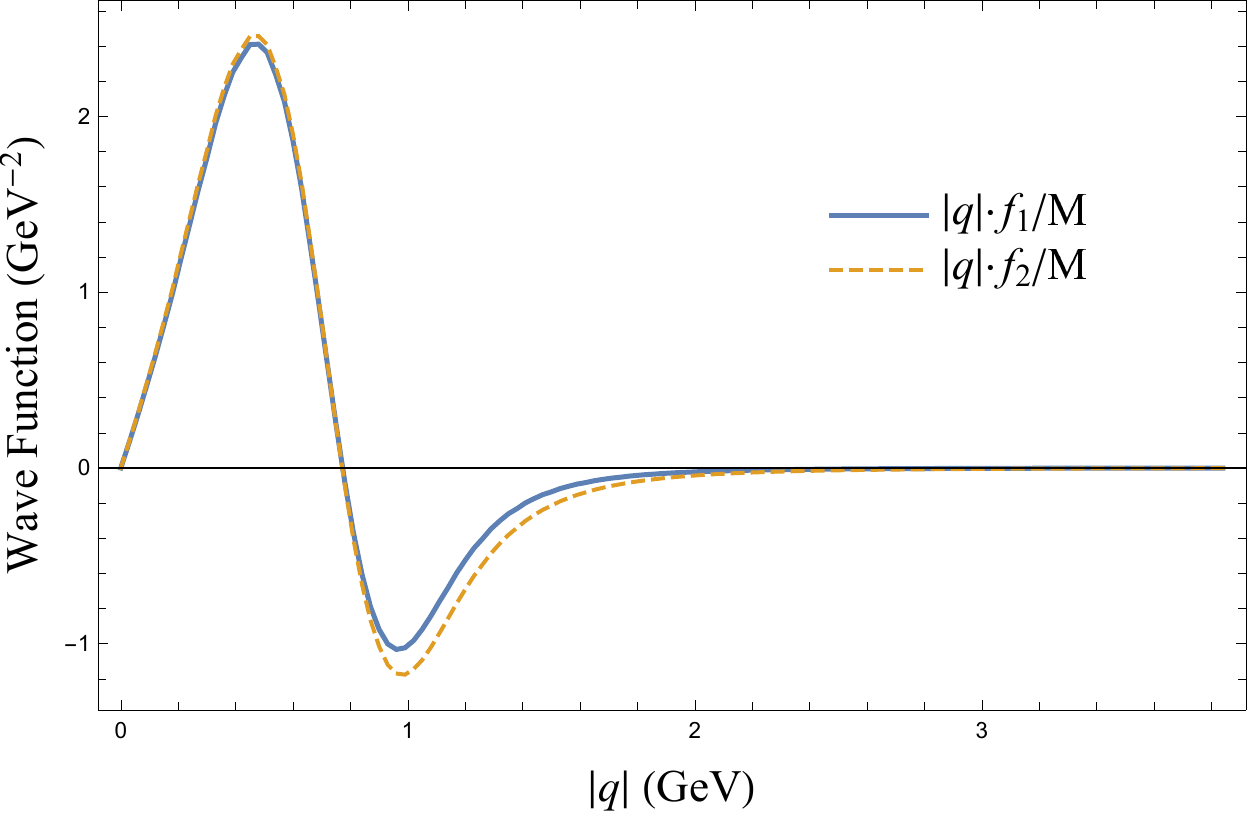}}
	\hspace{0.2cm}
	\subfigure[Wave functions of $ 3P $ state $ D_0^*(3P) $]{
		\label{3Pwf} %% label for first subfigure
		\includegraphics[width=0.45\textwidth]{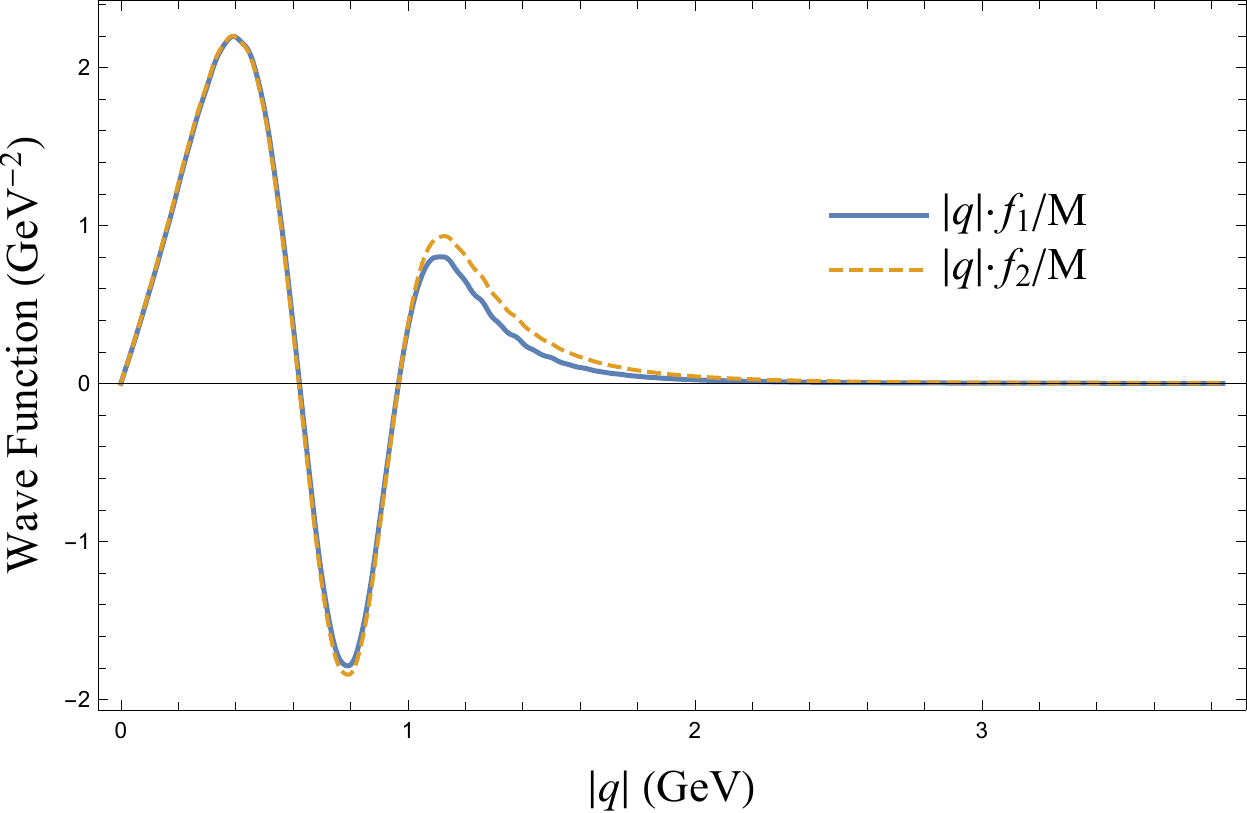}}
	\caption{Wave functions of $ B^- $ and $ D_0^* $.}
	\label{wfBD}
\end{figure}

Why the same parameters varying leads to the abnormal results of excited states? We consider that the different structures of BS wave function possibly play an important role here.   Fig. \ref{wfBD} shows the wave function values changing with the relative momentum$ |q| $. The wave functions of $ B^- $ and $ 1P $ state are all positive without nodes. When we varying the input parameters, the curve will have some small shift. And the shift could cause the small uncertainty in the overlapping integral. For excited states, we can find that the wave functions have nodes . For $ 2P $ states, the wave function changes from positive to negative after the nodes. In the overlapping integral, it causes the cancellation and the final results will be highly suppressed. Then if we vary the input parameters, a small shift of the wave function could cause a large uncertainty.

For $ 3P $ states, the wave function has two nodes, and the value change from negative to positive after the second node. The cancellation gets smaller than the case of the $ 2P $ state. Thus, the final branching ratio seems to be fine and the uncertainty is not very large.

\begin{figure}[htb!]
	\centering
	\subfigure[Branching ratio vers the mass of $D_0^*(2400), l=e$]{
		\label{br2400emu} %% label for first subfigure
		\includegraphics[width=0.45\textwidth]{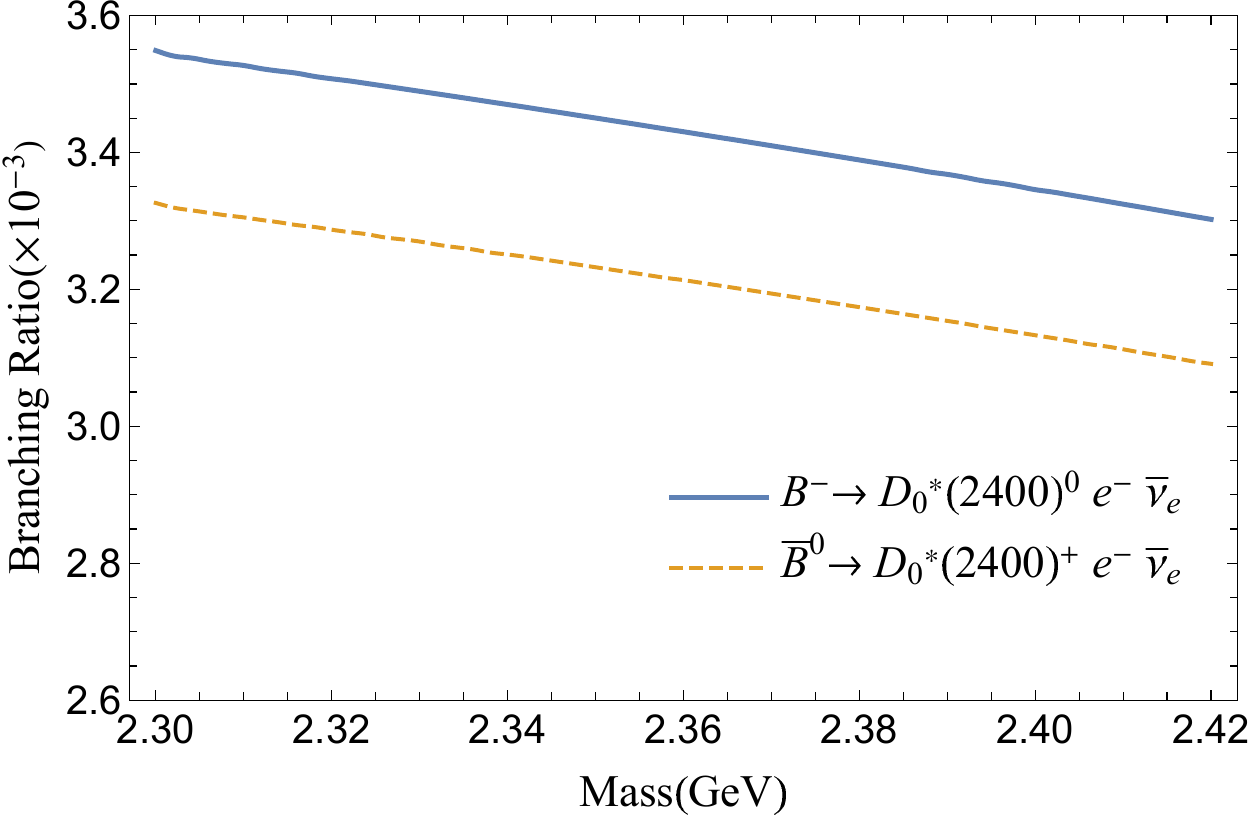}}
	\hspace{0.2cm}
	\subfigure[Braching ratio vers the mass of $D_0^*(2400), l=\tau$]{
		\label{br2400tau} %% label for first subfigure
		\includegraphics[width=0.45\textwidth]{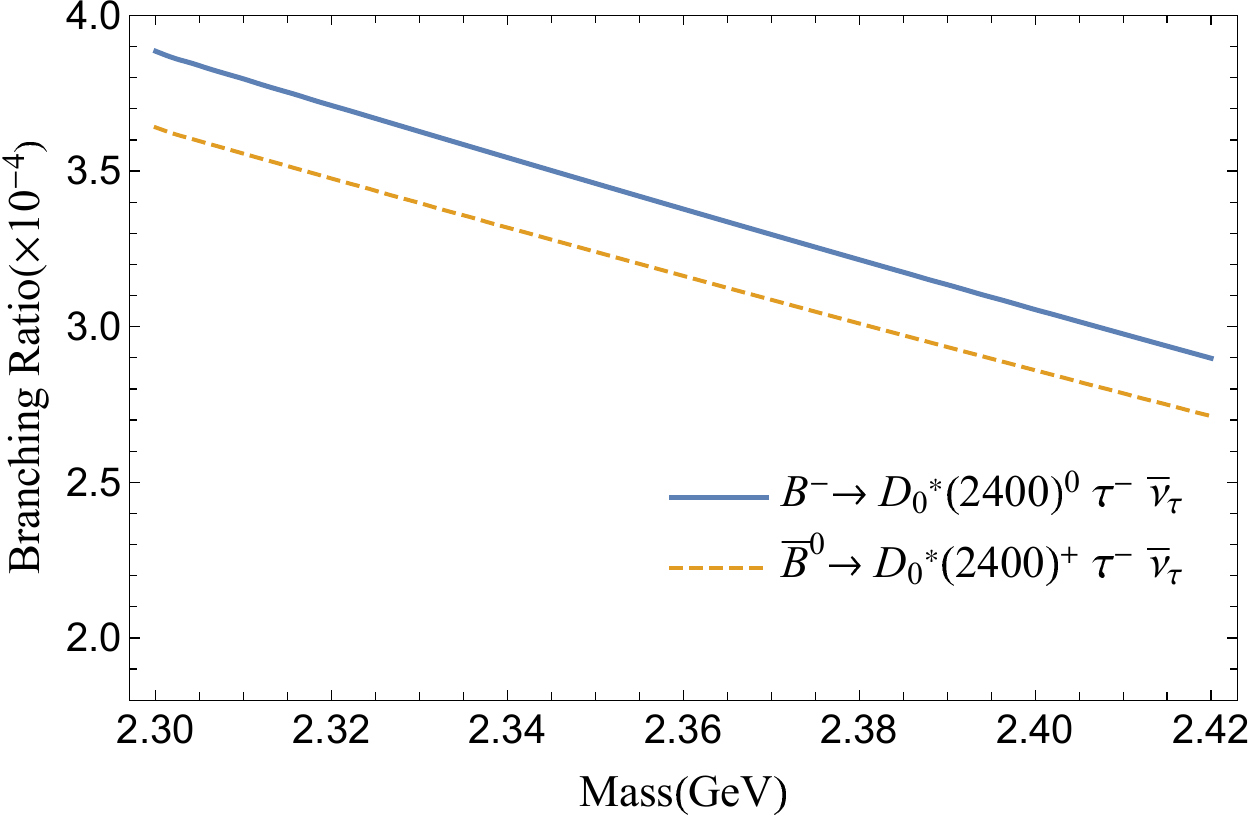}}

	\subfigure[Braching ratio vers the mass of $D_J^*(3000), l=e$]{
		\label{br3000emu} %% label for first subfigure
		\includegraphics[width=0.45\textwidth]{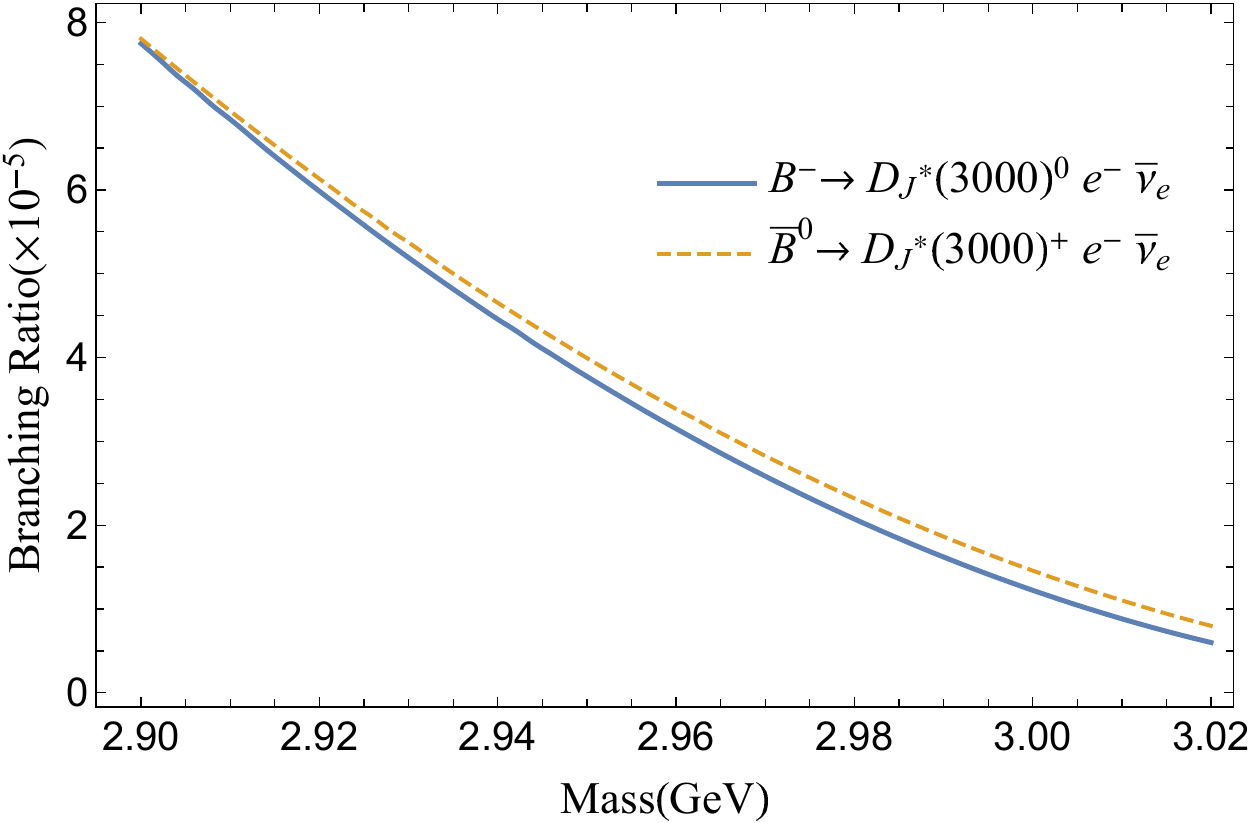}}
	\hspace{0.2cm}
	\subfigure[Braching ratio vers the mass of $D_J^*(3000), l=\tau$]{
		\label{br3000tau} %% label for first subfigure
		\includegraphics[width=0.45\textwidth]{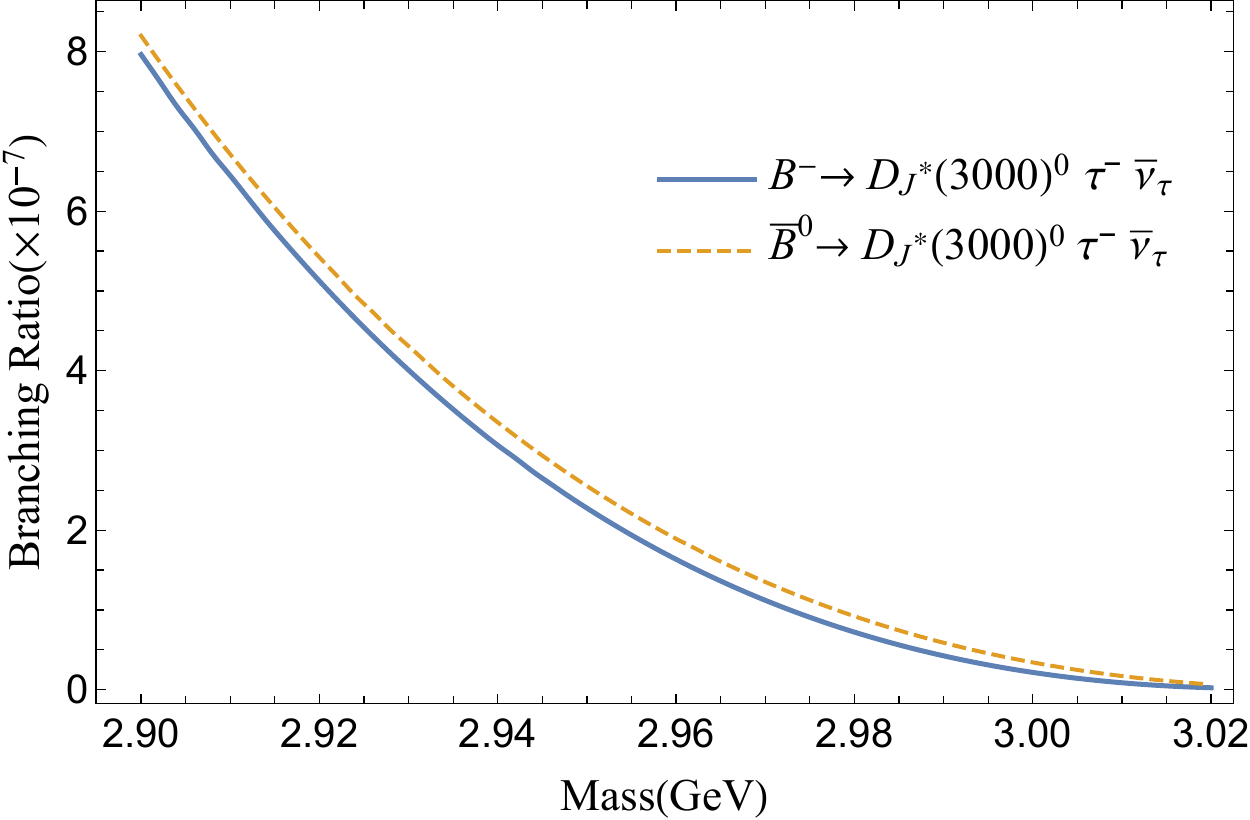}}

	\subfigure[Braching ratio vers the mass of $D_J^*(3P), l=e$]{
		\label{br3183emu} %% label for first subfigure
		\includegraphics[width=0.45\textwidth]{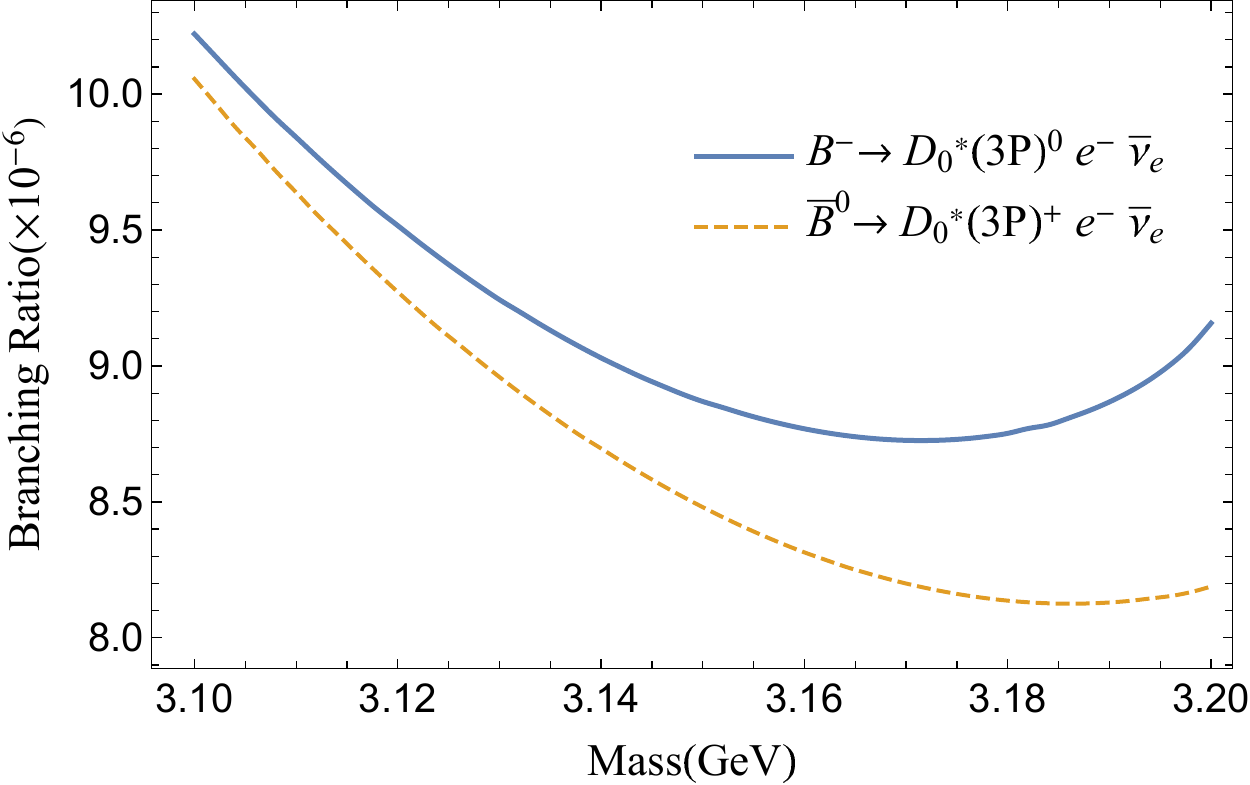}}
	\hspace{0.2cm}
	\subfigure[Braching ratio vers the mass of $D_J^*(3P), l=\tau$]{
		\label{br3183tau} %% label for first subfigure
		\includegraphics[width=0.45\textwidth]{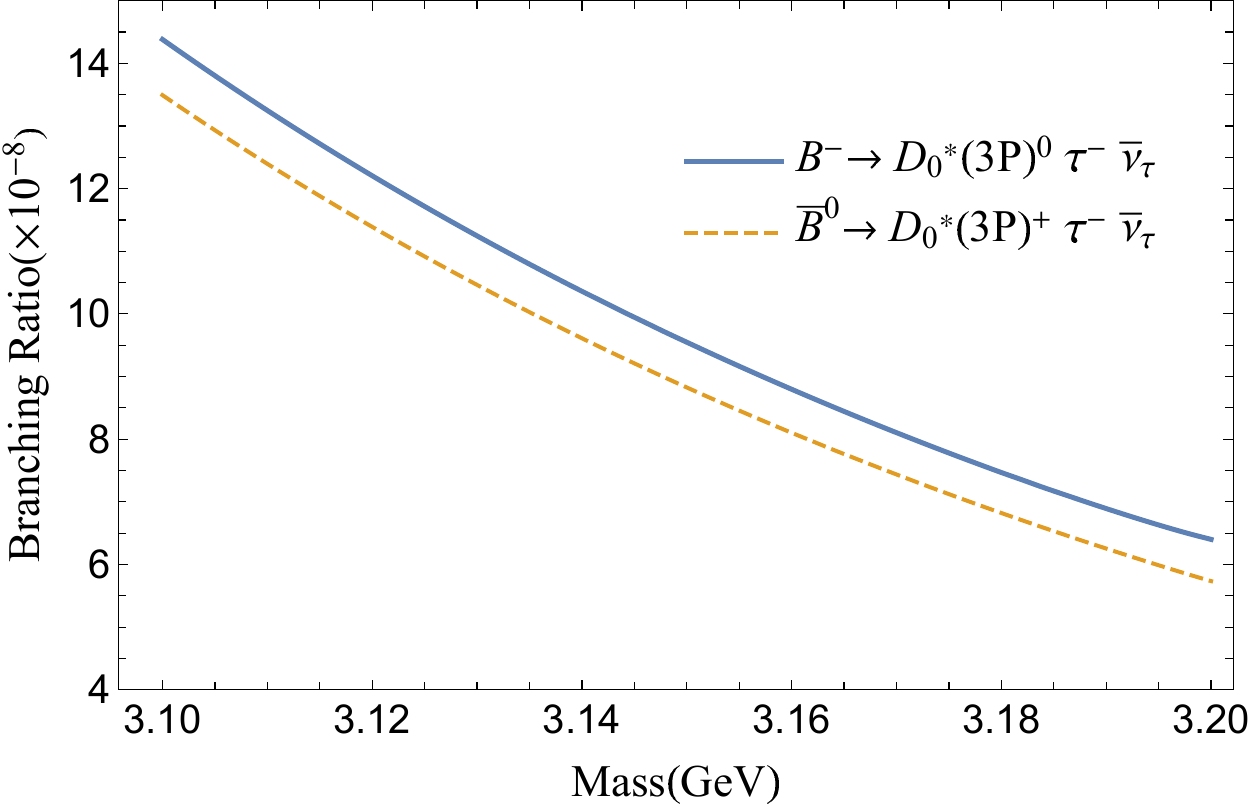}}
	\caption{The branching ratios of  semi-leptonic production change with the mass of $ D_0^* $.}
	\label{brmass}
\end{figure}

Considering the masses of these states have errors, the branching ratio of their semi-leptonic production changing with their masses are given in Fig. \ref{brmass}. For $ 1P $ state $ D_0^*(2400) $, the results have small changes. But the branching ratio of $ 2P $ state $ D_J^*(3000)$ dramatically decrease to nearly zero with increase of mass. The mass changing will also cause the wave function shift. That means the overlapping integral cancellation will increase as the mass increasing. For $ 3P $ state, it can be seen that the curves of $ l=e$ have minimum points around $ m=3.175 \GeV $, which means the overlapping integrals have the maximum cancellation at that mass value. After that, the values increase again. For $ l= \tau $, because of the small phase space, the branching ratios have the downtrend from the beginning to the end.

\begin{figure}[htb!]
	\centering
	\subfigure[Lepton spectra when $ l=e $]{
		\label{lse} %% label for first subfigure
		\includegraphics[width=0.45\textwidth]{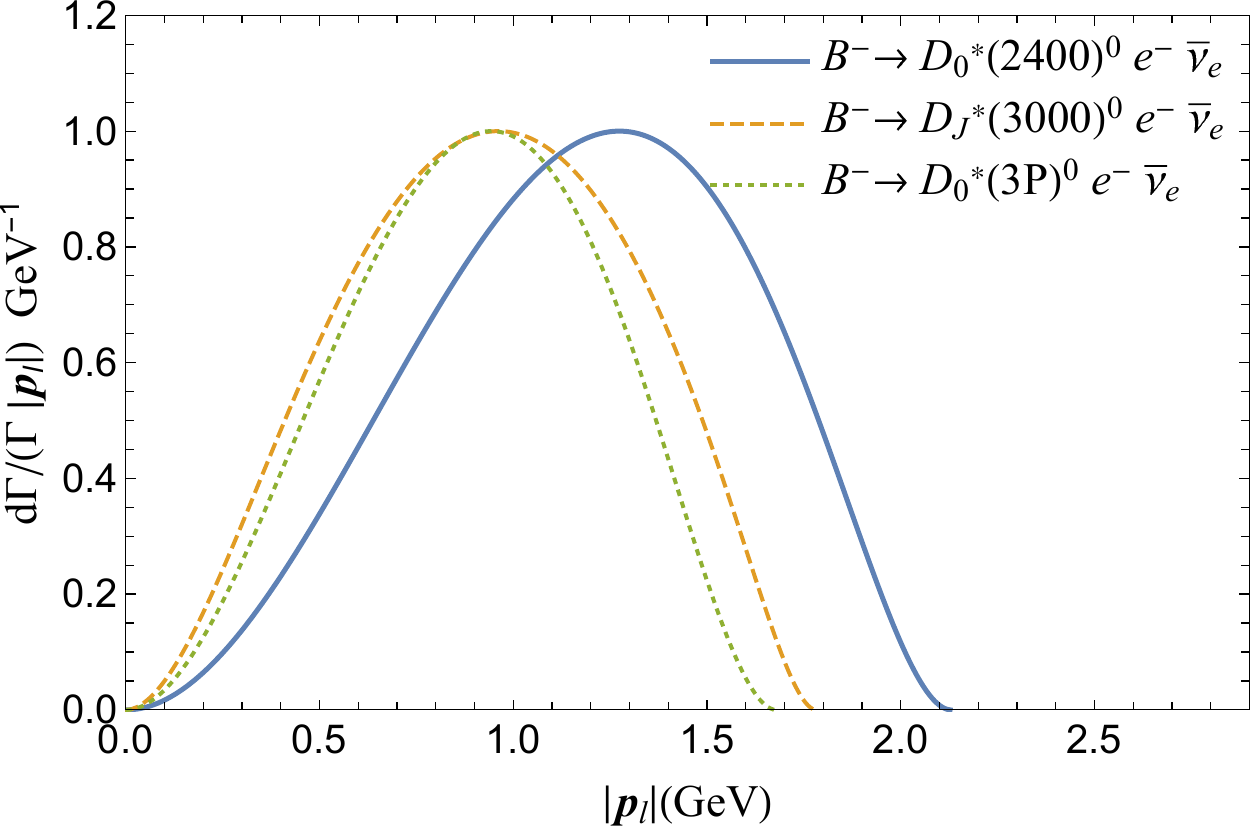}}
	\hspace{0.2cm}
	\subfigure[Lepton spectra when $ l=\tau $]{
		\label{lstau} %% label for first subfigure
		\includegraphics[width=0.45\textwidth]{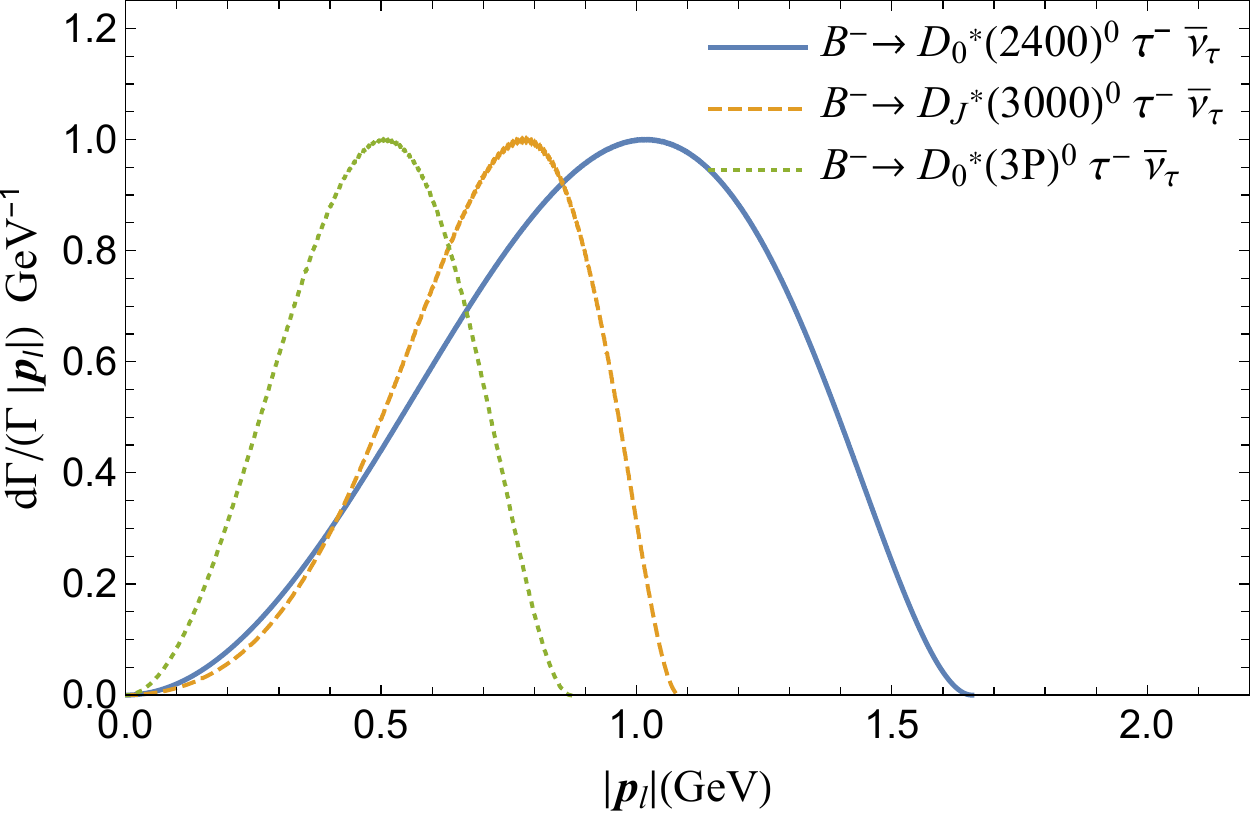}}
	\caption{The spectra of differential decay width vs $ |\bm{p}_l| $(with
		normalization).}
	\label{ls}
\end{figure}

Then, the normalized lepton spectra of the semi-leptonic production are presented in Fig. \ref{ls}. Because there are almost no difference between $ l=e $ and $ l=\mu $, only the channels of $ B^- \to D_0^{*0} e^- \overline{\nu}_e $ and $ B^- \to D_0^{*0} \tau^- \overline{\nu}_{\tau}  $ are given here. The spectrum peaks of $ 2P $ and $ 3P $ states move left because phase space decreases, especially for $ l=\tau $.

\subsection{Non-leptonic production}

The non-leptonic production results of $ 1P $ state $ D_0^*(2400) $ are shown in Table \ref{nonleptonic2400}. The $ D_0^*\pi $ and $ D_0^*\rho $ channels get the order of $ 10^{-3} $, while $ D_0^*K $ and $ D_0^*K^* $ modes are in the order of $ 10^{-5} $ and $ 10^{-4} $, respectively.

\begin{table}[htb!]
	\renewcommand\arraystretch{1.01}
	\caption[results]{Non-leptonic production results of the $ 1P $ state $ D_0^*(2400) $. }
	\label{nonleptonic2400}
	\vspace{0.5em}\centering
	\begin{tabular}{lcc}
		\toprule[1.5pt]
		\multicolumn{1}{c}{Channels} & Widths (GeV) & BR   \\
		\midrule[1pt]
		$ B^{-} \to D_0^*(2400)^0 \pi^-  $ & $ 3.35 \sim 5.78 \times 10^{-16} $ & $ 0.83 \sim 1.44 \times 10^{-3} $ \\
		 $ \overline{B}^{0} \to D_0^*(2400)^+ \pi^-  $ & $ 3.46 \sim 5.80 \times 10^{-16} $ & $ 0.80 \sim 1.34 \times 10^{-3} $  \\
		$ B^{-} \to D_0^*(2400)^0  K^- $ & $ 2.48 \sim 4.26 \times 10^{-17} $ & $ 0.62 \sim 1.06 \times 10^{-5} $ \\
		 $ \overline{B}^{0} \to D_0^*(2400)^+  K^- $ & $ 2.56 \sim 4.27 \times 10^{-17} $ & $ 5.91 \sim 9.85 \times 10^{-6} $ \\
		$ B^{-} \to D_0^*(2400)^0  \rho^- $ & $ 0.74 \sim 1.30\times10^{-15} $ & $ 1.85 \sim 3.23 \times 10^{-3} $ \\
		 $ \overline{B}^{0} \to D_0^*(2400)^+  \rho^- $ & $ 0.76 \sim 1.30 \times10^{-15} $ & $ 1.77 \sim 2.99 \times 10^{-3} $ \\
		$ B^{-} \to D_0^*(2400)^0 K^{*-}$ & $ 4.25 \sim 7.46 \times 10^{-17} $ & $ 1.06 \sim 1.86 \times 10^{-4} $ \\
		 $ \overline{B}^{0} \to D_0^*(2400)^+ K^{*-}$ & $ 4.37 \sim 7.44 \times 10^{-17} $ & $ 1.01 \sim 1.72 \times 10^{-4} $ \\
		\bottomrule[1.5pt]
	\end{tabular}
\end{table}

There have been some experimental results of their cascade decays. For comparison, our results of the cascade decays are shown in Table \ref{cascadenl}.  The Belle and BABAR Collaborations show that the branching ratios $ \mathcal{B}\left(B^{-} \rightarrow D_{0}^{* 0} \pi^{-}\right) \times \mathcal{B}\left(D_{0}^{* 0} \rightarrow D^{+} \pi^{-}\right)$ are $ (6.1 \pm 0.6 \pm 0.9 \pm 1.6) \times 10^{-4} $\cite{Belle2004} and $ (6.8 \pm 0.3 \pm 0.4 \pm 2.0) \times 10^{-4} $\cite{BABAR2009.112004}, respectively. Our results are consistent with them. For the charged $ D_0^{*\pm} $ meson, the Belle and LHCb Collaborations give the results $ \mathcal{B}(\overline{B}^{0} \rightarrow D_{0}^{*+} \pi^{-}) \times \mathcal{B}\left(D_{0}^{*+} \rightarrow D^{0} \pi^{+}\right) =(6.0 \pm 1.3 \pm 1.5 \pm 2.2) \times 10^{-5}$\cite{Belle2007.012006} and $ (7.7 \pm 0.5 \pm 0.3 \pm 0.3 \pm 0.4) \times 10^{-5}$\cite{LHCb2015.032002}, respectively, which are much lower than the branching ratio of the previous channel. For $ D_0^*K $ channel, the LHCb collaboration shows the result  $ \mathcal{B}\left(B^{-} \rightarrow D_0^{*0} K^{-}\right) \times \mathcal{B}\left(D_0^{*0} \rightarrow D^{+} \pi^{-}\right)=(6.1 \pm 1.9 \pm 0.5 \pm 1.4 \pm 0.4) \times 10^{-6} $\cite{LHCb2016.119901}, which is also lower than our calculation. From the perspective of symmetry, the non-leptonic results of these channels should be similar. But different experimental results show marked discrepancy, which need more experimental data accumulations and theoretical attentions.

\begin{table}[htb!]
	\renewcommand\arraystretch{0.90}
	\caption[results]{The branching ratios of cascade decays to $ D \pi $. }
	\label{cascadenl}
	\vspace{0.5em}\centering
	\resizebox{\textwidth}{!}{
		\begin{tabular}{llcllc}
			\toprule[1.5pt]
			\multicolumn{2}{c}{Channels} & BR & \multicolumn{2}{c}{Channels} & BR \\
			\midrule[1pt]
			\multirow{2}{*}{$ B^{-} \to D_0^*(2400)^0 \pi^- $}	& $  \to D^0 \pi^0 $ &  $ 2.76 \sim 4.76 \times 10^{-4} $ & \multirow{2}{*}{$ \overline{B}^{0} \to D_0^*(2400)^+ \pi^- $}	& $  \to D^0 \pi^+ $ &  $ 5.34 \sim 8.94 \times 10^{-4} $   \\
			& $ \to D^+ \pi^- $ &  $ 5.57 \sim 9.63  \times 10^{-4} $ & & $ \to D^+ \pi^0 $ &  $ 2.66 \sim 4.45  \times 10^{-4} $  \\
			
			\multirow{2}{*}{$ B^{-} \to D_0^*(2400)^0 K^- $}	& $  \to D^0 \pi^0 $ &  $ 2.04 \sim 3.51\times 10^{-5} $ & \multirow{2}{*}{$ \overline{B}^{0} \to D_0^*(2400)^+ K^- $}	& $  \to D^0 \pi^+ $ &  $ 3.95 \sim 6.58 \times 10^{-5} $   \\
			& $ \to D^+ \pi^- $ &  $ 4.12 \sim 7.09  \times 10^{-5} $ & & $ \to D^+ \pi^0 $ &  $ 1.96 \sim 3.27  \times 10^{-5} $  \\
			
			\multirow{2}{*}{$ B^{-} \to D_0^*(2400)^0 \rho^- $}	& $  \to D^0 \pi^0 $ &  $ 0.61 \sim 1.07\times 10^{-3} $ & \multirow{2}{*}{$ \overline{B}^{0} \to D_0^*(2400)^+ \rho^- $}	& $  \to D^0 \pi^+ $ &  $ 1.18 \sim 2.00 \times 10^{-3} $   \\
			& $ \to D^+ \pi^- $ &  $ 1.23 \sim 2.16  \times 10^{-3} $ & & $ \to D^+ \pi^0 $ &  $ 5.86 \sim 9.93  \times 10^{-4} $  \\
			
			\multirow{2}{*}{$ B^{-} \to D_0^*(2400)^0 K^{*-} $}	& $  \to D^0 \pi^0 $ &  $ 3.50 \sim 6.15 \times 10^{-5} $ & \multirow{2}{*}{$ \overline{B}^{0} \to D_0^*(2400)^+ K^{*-} $}	& $  \to D^0 \pi^+ $ &  $ 0.68 \sim 1.15 \times 10^{-4} $   \\
			& $ \to D^+ \pi^- $ &  $ 0.71 \sim 1.24  \times 10^{-4} $ & & $ \to D^+ \pi^0 $ &  $ 3.35 \sim 5.71  \times 10^{-5} $  \\
			
			\bottomrule[1.5pt]
	\end{tabular}}
\end{table}

Then, like the previous section, the non-leptonic productions of $ 2P $ and  $ 3P $ states are also considered and the results are shown in Table \ref{nonleptonic3000} and \ref{nonleptonic3183}, respectively.

\begin{table}[htb!]
	\renewcommand\arraystretch{1.01}
	\caption[results]{Non-leptonic production results of the $ 2P $ state $ D_0^*(3000) $. }
	\label{nonleptonic3000}
	\vspace{0.5em}\centering
	\begin{tabular}{lcc}
		\toprule[1.5pt]
		\multicolumn{1}{c}{Channels} & Widths (GeV) & BR  \\
		\midrule[1pt]
		 $ B^{-} \to D_0^*(3000)^0 \pi^-  $ & $ 0.07\sim 54.7 \times 10^{-18} $ & $ 0.02\sim 13.4 \times 10^{-5} $ \\ $ \overline{B}^{0} \to D_0^*(3000)^+ \pi^-  $ & $ 0.003\sim 55.3 \times 10^{-18} $ & $ 0.0006\sim 12.8 \times 10^{-5} $ \\
		 $ B^{-} \to D_0^*(3000)^0  K^- $ & $ 0.08\sim 37.9\times 10^{-19} $ & $0.02\sim 9.44 \times 10^{-6} $ \\ $ \overline{B}^{0} \to D_0^*(3000)^+  K^- $ & $ 0.02\sim 39.0\times 10^{-19} $ & $ 0.004\sim 9.02 \times 10^{-6} $ \\
		 $ B^{-} \to D_0^*(3000)^0  \rho^- $ & $ 0.04\sim 10.8\times10^{-17} $ & $ 0.10\sim 26.8\times 10^{-5} $ \\ $ \overline{B}^{0} \to D_0^*(3000)^+  \rho^- $ & $ 0.01\sim 11.1\times10^{-17} $ & $ 0.03\sim 25.7 \times 10^{-5} $ \\
		 $ B^{-} \to D_0^*(3000)^0 K^{*-}$ & $ 0.28\sim 59.0\times 10^{-19} $ & $ 0.07\sim 14.7 \times 10^{-6} $ \\ $ \overline{B}^{0} \to D_0^*(3000)^+ K^{*-}$ & $ 0.11\sim 61.2 \times 10^{-19} $ & $ 0.03\sim 14.1 \times 10^{-6} $ \\
		\bottomrule[1.5pt]
	\end{tabular}
\end{table}

Similar to the semi-leptonic occasion, the results of $ 2P $ state $ D_0^*(3000) $ have a large uncertainty , which can also be explained by the node structure of BS wave function. Because the phase spaces of $ 2P $ and $ 3P $ state are close, their branching ratios reach similar magnitude.

\begin{table}[htb!]
	\renewcommand\arraystretch{1.01}
	\caption[results]{Non-leptonic production results of the $ 3P $ state $ D_0^* $. }
	\label{nonleptonic3183}
	\vspace{0.5em}\centering
		\begin{tabular}{lcc}
			\toprule[1.5pt]
			\multicolumn{1}{c}{Channels} & Widths (GeV) & BR   \\
			\midrule[1pt]
			$ B^{-} \to D_0^*(3P)^0 \pi^-  $ & $ 1.87 \sim 5.59 \times 10^{-18} $ & $ 0.47 \sim 1.39 \times 10^{-5} $ \\ $ \overline{B}^{0} \to D_0^*(3P)^+ \pi^-  $ & $ 1.86 \sim 6.00 \times 10^{-18} $ & $ 0.43 \sim 1.39 \times 10^{-5} $  \\
			$ B^{-} \to D_0^*(3P)^0  K^- $ & $ 1.43 \sim 4.02 \times 10^{-19} $ & $ 0.36 \sim 1.00 \times 10^{-6} $ \\ $ \overline{B}^{0} \to D_0^*(3P)^+  K^- $ & $ 1.41 \sim 4.30 \times 10^{-19} $ & $ 0.33 \sim 0.99 \times 10^{-6} $ \\
			$ B^{-} \to D_0^*(3P)^0  \rho^- $ & $ 0.38 \sim 1.03 \times10^{-17} $ & $ 0.96 \sim 2.56 \times 10^{-5} $ \\ $ \overline{B}^{0} \to D_0^*(3P)^+  \rho^- $ & $ 0.38 \sim 1.11 \times10^{-17} $ & $ 0.89 \sim 2.55 \times 10^{-5} $ \\
			$ B^{-} \to D_0^*(3P)^0 K^{*-}$ & $ 2.10 \sim 5.47  \times 10^{-19} $ & $ 0.52 \sim 1.36 \times 10^{-6} $ \\ $ \overline{B}^{0} \to D_0^*(3P)^+ K^{*-}$ & $ 2.11 \sim 5.89 \times 10^{-19} $ & $ 0.49 \sim 1.36 \times 10^{-6} $ \\
			\bottomrule[1.5pt]
	\end{tabular}
\end{table}

\begin{figure}[htb!]]
	\centering
	\subfigure[Braching ratio vers the mass of $ 1P $ state $D_0^*(2400)^0$]{
		\label{nl2400zero} %% label for first subfigure
		\includegraphics[width=0.45\textwidth]{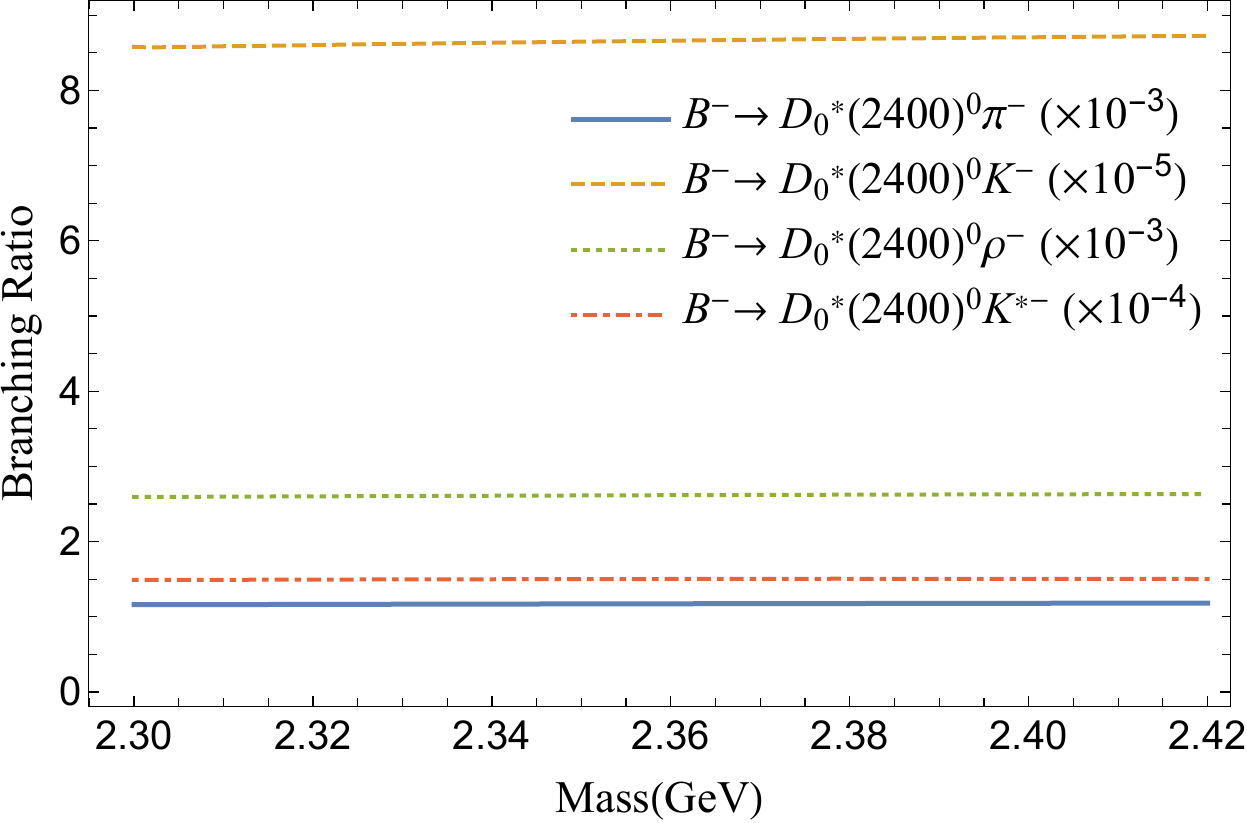}}
	\hspace{0.2cm}
	\subfigure[Braching ratio vers the mass of $ 1P $ state $D_0^*(2400)^+$]{
		\label{nl2400plusl} %% label for first subfigure
		\includegraphics[width=0.45\textwidth]{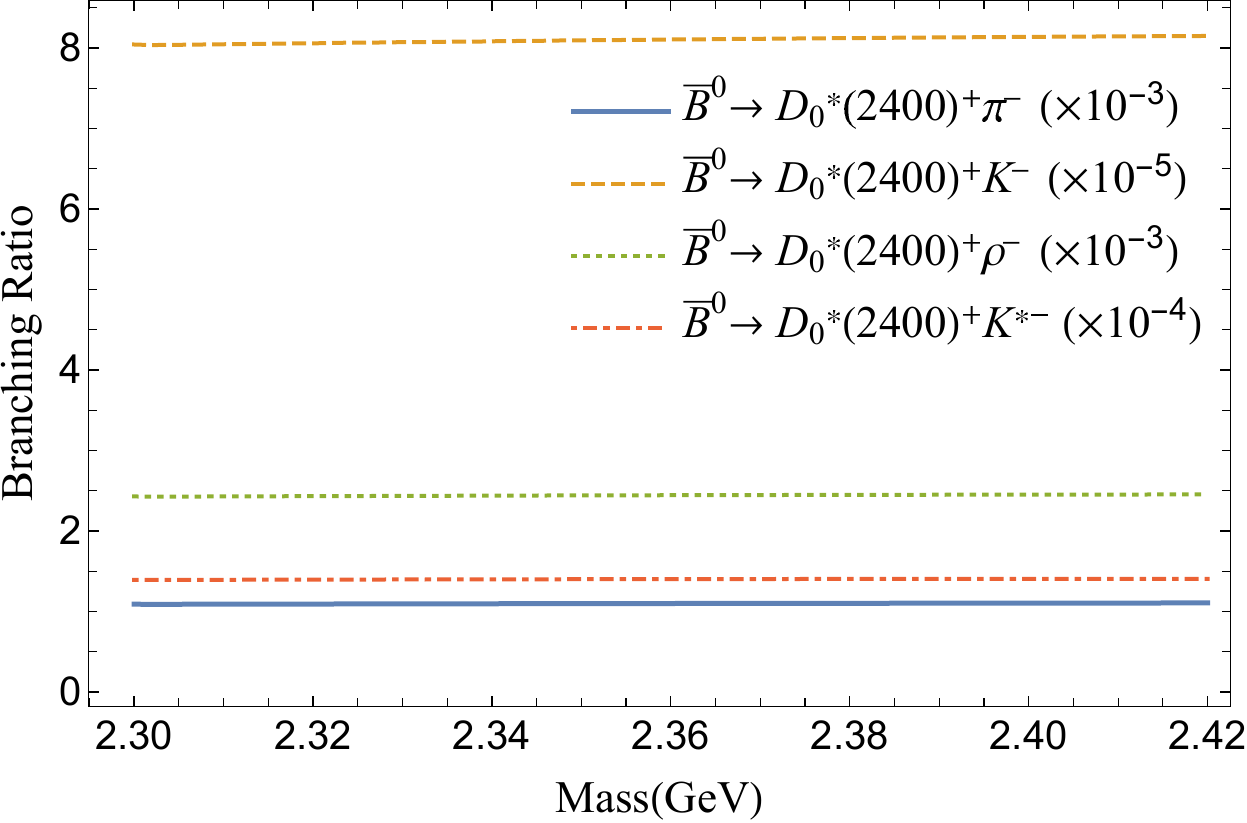}}
	
	\subfigure[Braching ratio vers the mass of $ 2P $ state $D_J^*(3000)^0$]{
		\label{nl3000zero} %% label for first subfigure
		\includegraphics[width=0.45\textwidth]{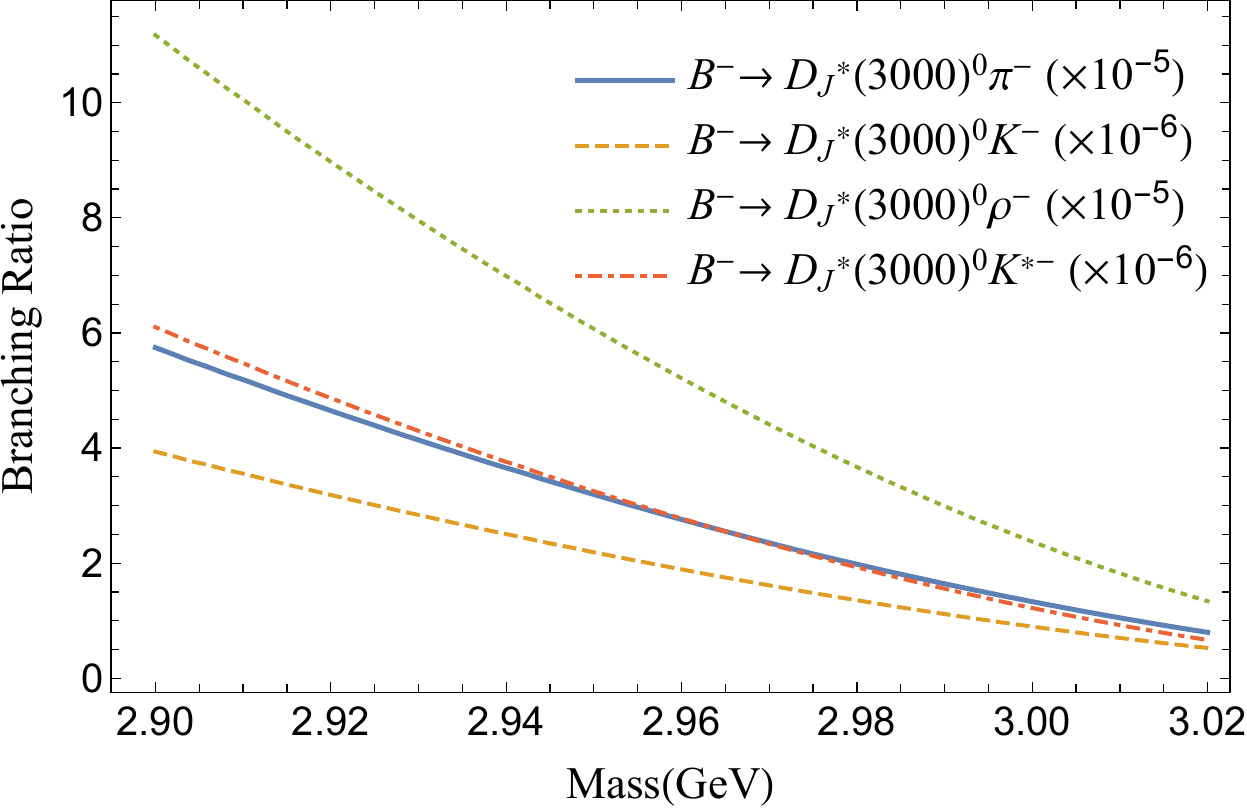}}
	\hspace{0.2cm}
	\subfigure[Braching ratio vers the mass of $ 2P $ state $D_J^*(3000)^+$]{
		\label{nl3000plus} %% label for first subfigure
		\includegraphics[width=0.45\textwidth]{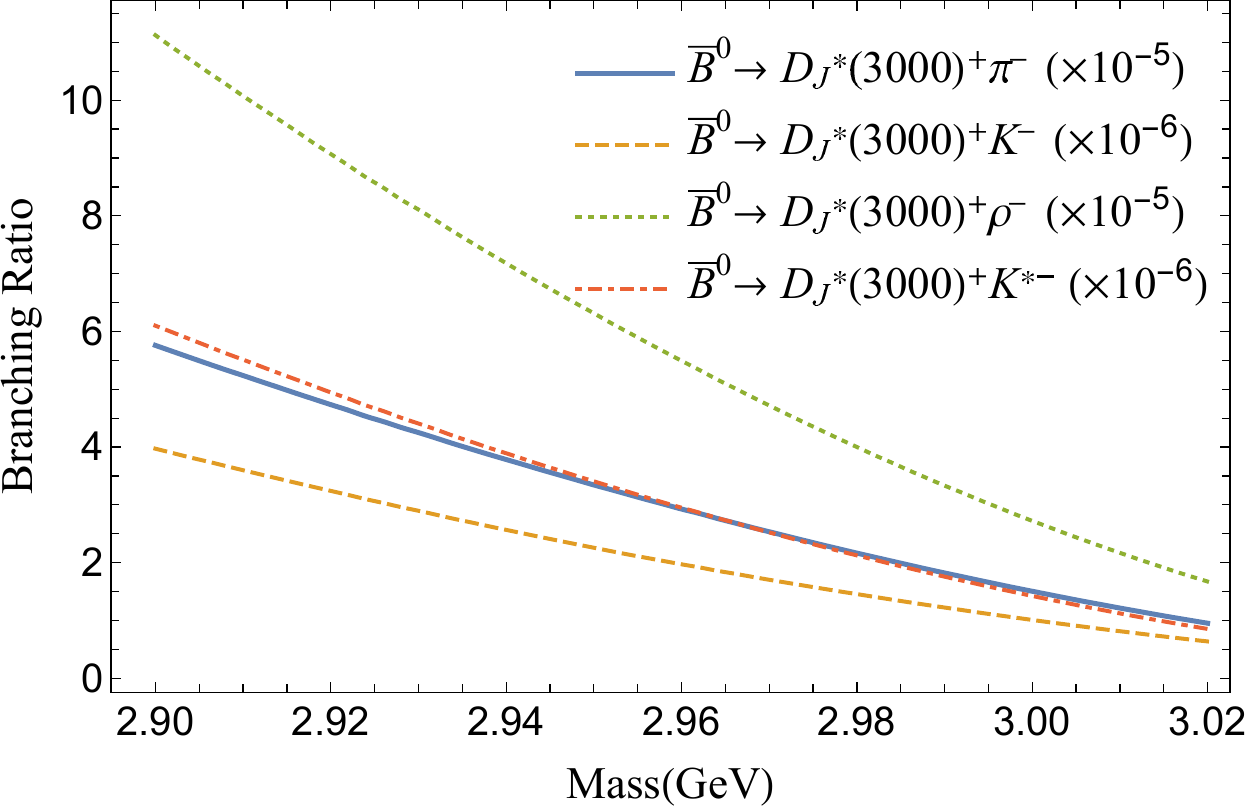}}
	
	\subfigure[Braching ratio vers the mass of $ 3P $ state $D_0^*(3183)^0$]{
		\label{nl33183zero} %% label for first subfigure
		\includegraphics[width=0.45\textwidth]{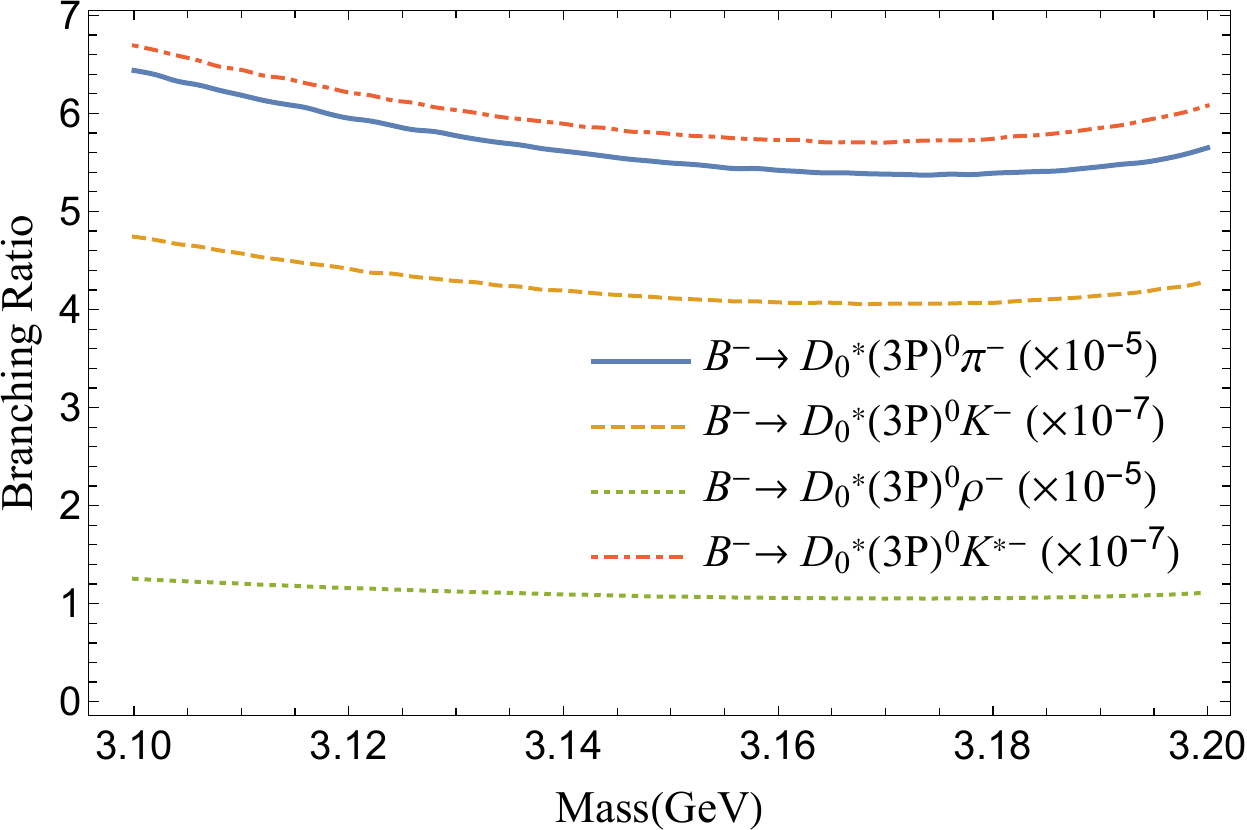}}
	\hspace{0.2cm}
	\subfigure[Braching ratio vers the mass of $ 3P $ state $D_0^*(3183)^+$]{
		\label{nl3183plus} %% label for first subfigure
		\includegraphics[width=0.45\textwidth]{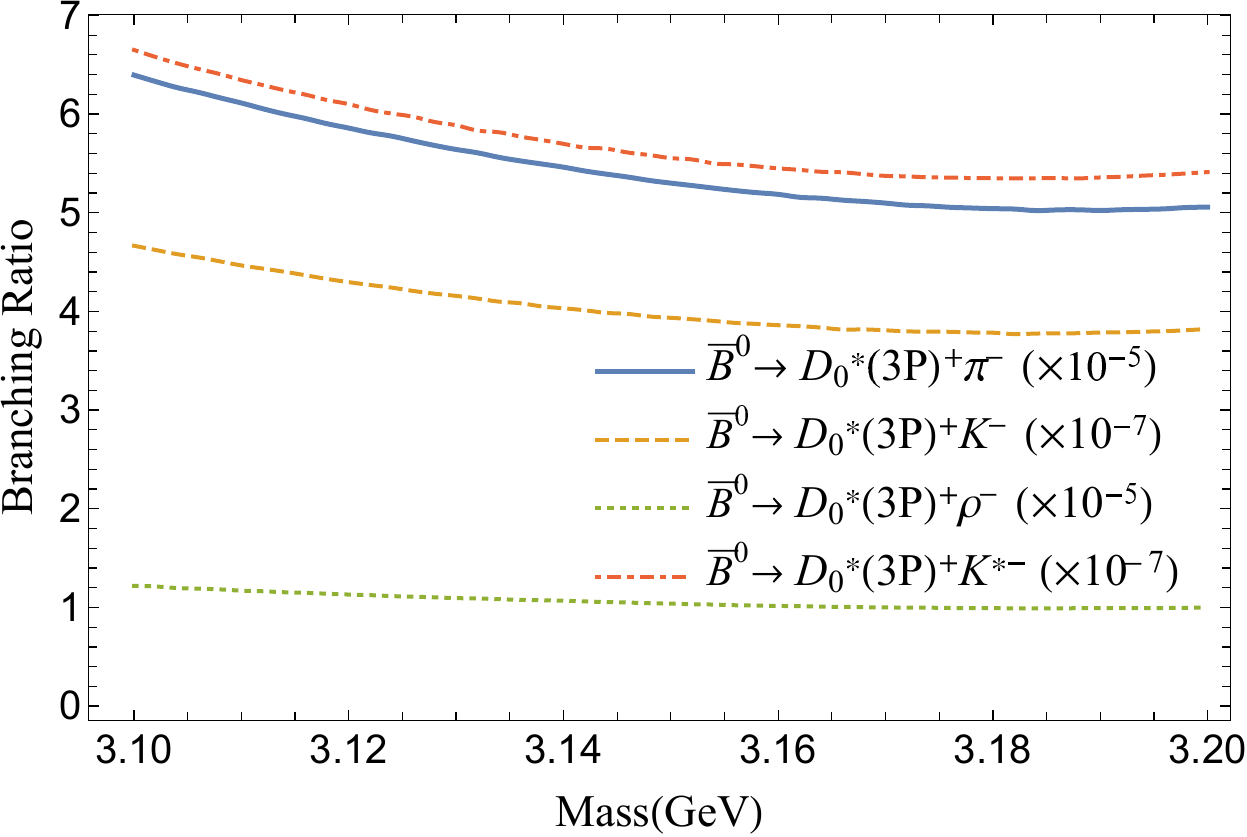}}
	\caption{The branching ratios of   non-leptonic production of $ D_0^* $.}
	\label{nonleptonic}
\end{figure}

We also draw the branching ratios changing with the mass of $ D_0^* $ in Fig. \ref{nonleptonic}. Like the semi-leptonic production case, the non-leptonic production results of  $ 2P $ state are sensitive to the mass. The curves of $ 1P $ and $ 3P $ states stay relatively stable when the mass values change. There are also minimum points of $ 3P $ states' curves at $ 3.175\GeV $, where the maximum cancellation of overlapping integral occurs.

\section{SUMMARY}

Based on the instantaneous BS framework, we calculate the semi-leptonic and non-leptonic productions of several excited $ D_0^* $ states from $ B$ mesons. For $ 1P $ state $ D_0^*(2400) $, the branching ratios of $ B \to D_0^* e^- \overline{\nu}_{e} $ are in the order of $ 10^{-3} $, which is consistent with the results of present experiments and other models. For non-leptonic channels, the experiments didn't get quite consistent results while our calculating consists with parts of present experimental results.
For $ 2P $ states $ D_J^*(3000) $, we get suppressed branching ratios in the order of $ 10^{-5} \sim 10^{-6}$ and large uncertainty in both semi-leptonic and non-leptonic channels. The cancellation in overlapping integral, which is caused by its one-nodes structure of the BS wave functions, could explain the abnormal results. For $ 3P $ states, their ratios are of the same order of magnitude as $ 2P $ states' results because they have similar phase spaces. The two-nodes structure of $ 3P $ states wave functions makes the cancellation smaller than that of $ 2P $ states and get the minimum branching ratios if their masses are around 3.175 GeV.
Our work could give some inspiration to future experiment and we expect more attention on these production processes of the orbitally excited $ D $ mesons.

%%%%%%%%%%%%%%%%%%%%%%%%%%%%%%%%%%%%%%%%%%%%%%%%%%%%%%%%%%%%%%%%

\section*{ACKNOWLEDGEMENTS}
This work was supported in part by the National Natural Science
Foundation of China (NSFC) under Grant No.~11575048, No.~11405037, No.~11505039.
We thank the HPC Studio at Physics Department of Harbin Institute of Technology for access to computing resources through INSPUR-HPC@PHY.HIT.

%%%%%%%%%%%%%%%%%%%%%%%%%%%%%%%%%%%%%%%%%%%%%%%%%%%%%%%%%%%%%%%%

\appendix
\section*{APPENDIX  Bethe-Salpeter Wave Function}
\setcounter{equation}{0}

The general forms of wave functions are 
\begin{equation}
\varphi_{0^+}(q_{\perp})=M \left[
\frac{\slashed{q}_{\perp}}{M}f_{a1}(q_{\perp})+\frac{\slashed{P}\slashed{q}_{\perp}}{M^2} f_{a2}(q_{\perp})+f_{a3}(q_{\perp})+\frac{\slashed{P}}{M}f_{a4}(q_{\perp})
\right],
\end{equation}
\begin{equation}
\varphi_{0^-}(q_{\perp})=M\left[
\frac{\slashed{P}}{M}f_{b1}(q_\perp)+f_{b2}(q_\perp)+\frac{\slashed{q}_\perp}{M}f_{b3}(q_\perp)+\frac{\slashed{P}\slashed{q}_\perp}{M^2}f_{b4}(q_\perp)
\right] \gamma_5,
\end{equation}
where the constraint conditions are 
\begin{equation}
\begin{split}
f_{a3}=\frac{q_{\perp}^2(\omega_1+\omega_2)}{M(m_1 \omega_2+m_2 \omega_1)}f_{a1},\  f_{a4}=\frac{q_{\perp}^2(\omega_1-\omega_2)}{M(m_1\omega_2+m_2\omega_1)}f_{a2},
\end{split}
\end{equation}

\begin{equation}
\begin{split}
f_{b3}=\frac{M(\omega_2-\omega_1)}{m_1\omega_2+m_2\omega_1} f_{b2},\  f_{b4}=-\frac{M(\omega_1+\omega_2}{m_1\omega_2+m_2\omega_1}f_{b1}.
\end{split}
\end{equation}

The positive parts are expressed as 
\begin{equation}\label{0+}
\varphi_{0^+}^{++}(q_{\perp})=A_1+A_2\frac{\slashed{P}}{M}+A_3\frac{\slashed{q}_{\perp}}{M}+A_4\frac{\slashed{P}\slashed{q}_{\perp}}{M^2},
\end{equation}
\begin{equation}
\varphi_{0^-}^{++}(q_{\perp}) = \left[ B_1(q_{\perp})+\frac{\slashed{P}}{M}B_2(q_{\perp})+\frac{\slashed{P} }{M}B_3(q_{\perp})+\frac{\slashed{P} \slashed{q}_{\perp}}{M^2}B_4(q_{\perp})
\right]\gamma _5,
\end{equation}

where
\begin{equation}
\begin{split}
A_1=\frac{(\omega_1+\omega_2)q^2_{\perp}}{2(m_1\omega_2+m_2\omega_1)}\left(
f_{a1}+\frac{m_1+m_2}{\omega _1+\omega_2}f_{a2}
\right),& \ 
A_3=\frac{M}{2}\left(
f_{a1}+\frac{m_1+m_2}{\omega_1+\omega_2}f_{a2}
\right),\\
A_2= \frac{(m_1-m_2)q^2_{\perp}}{2(m_1\omega_2+m_2\omega_1)}\left(
f_{a1}+\frac{m_1+m_2}{\omega _1+\omega_2}f_{a2}
\right), & \ 
A_4=\frac{M}{2}\left(
\frac{\omega_1+\omega_2}{m_1+m_2}f_{a1}+f_{a2}
\right),
\end{split}
\end{equation}
\begin{equation}
\begin{split}
&B_1 =\frac{M}{2}\left(
\frac{\omega _1+\omega _2}{m_1+m_2}f_{b1}+f_{b2}
\right),\  B_3= -\frac{M(\omega _1 -\omega _2)}{m_1 \omega _2 +m_2 \omega _1}B_1, \\
& B_2 =\frac{M}{2}\left(
f_{b1}+\frac{m_1+m_2}{\omega _1 +\omega _2}f_{b2}
\right), \ B_4 =- \frac{(m_1+m_2)M}{m_1 \omega _2+m_2 \omega _1}B_1.
\end{split}
\end{equation}

%%%%%%%%%%%%%%%%%%%%%%%%%%%%%%%%%%%%%%%%%%%%%%%%%%%%%%%%%%%%%%%%%%%%%%%%%%%%%%%%%%%%%%%%

%%%%%%%%%%%%%%%%%references%%%%%%%%%%%%%%%%%%%%%%%%%%%
\bibliography{paper.bib}

\providecommand{\noopsort}[1]{}\providecommand{\singleletter}[1]{#1}%
\begin{thebibliography}{38}%
\makeatletter
\providecommand \@ifxundefined [1]{%
 \@ifx{#1\undefined}
}%
\providecommand \@ifnum [1]{%
 \ifnum #1\expandafter \@firstoftwo
 \else \expandafter \@secondoftwo
 \fi
}%
\providecommand \@ifx [1]{%
 \ifx #1\expandafter \@firstoftwo
 \else \expandafter \@secondoftwo
 \fi
}%
\providecommand \natexlab [1]{#1}%
\providecommand \enquote  [1]{``#1''}%
\providecommand \bibnamefont  [1]{#1}%
\providecommand \bibfnamefont [1]{#1}%
\providecommand \citenamefont [1]{#1}%
\providecommand \href@noop [0]{\@secondoftwo}%
\providecommand \href [0]{\begingroup \@sanitize@url \@href}%
\providecommand \@href[1]{\@@startlink{#1}\@@href}%
\providecommand \@@href[1]{\endgroup#1\@@endlink}%
\providecommand \@sanitize@url [0]{\catcode `\\12\catcode `\$12\catcode
  `\&12\catcode `\#12\catcode `\^12\catcode `\_12\catcode `\%12\relax}%
\providecommand \@@startlink[1]{}%
\providecommand \@@endlink[0]{}%
\providecommand \url  [0]{\begingroup\@sanitize@url \@url }%
\providecommand \@url [1]{\endgroup\@href {#1}{\urlprefix }}%
\providecommand \urlprefix  [0]{URL }%
\providecommand \Eprint [0]{\href }%
\providecommand \doibase [0]{http://dx.doi.org/}%
\providecommand \selectlanguage [0]{\@gobble}%
\providecommand \bibinfo  [0]{\@secondoftwo}%
\providecommand \bibfield  [0]{\@secondoftwo}%
\providecommand \translation [1]{[#1]}%
\providecommand \BibitemOpen [0]{}%
\providecommand \bibitemStop [0]{}%
\providecommand \bibitemNoStop [0]{.\EOS\space}%
\providecommand \EOS [0]{\spacefactor3000\relax}%
\providecommand \BibitemShut  [1]{\csname bibitem#1\endcsname}%
\let\auto@bib@innerbib\@empty
%</preamble>
\bibitem [{\citenamefont {Buskulic}\ \emph {et~al.}(1997)\citenamefont
  {Buskulic} \emph {et~al.}}]{ALEPH1997.373}%
  \BibitemOpen
  \bibfield  {author} {\bibinfo {author} {\bibfnamefont {D.}~\bibnamefont
  {Buskulic}} \emph {et~al.} (\bibinfo {collaboration} {ALEPH Collaboration}),\
  }\href {\doibase 10.1016/S0370-2693(97)00071-3} {\bibfield  {journal}
  {\bibinfo  {journal} {Physics Letters B}\ }\textbf {\bibinfo {volume}
  {395}},\ \bibinfo {pages} {373} (\bibinfo {year} {1997})}\BibitemShut
  {NoStop}%
\bibitem [{\citenamefont {Bartelt}\ \emph {et~al.}(1999)\citenamefont {Bartelt}
  \emph {et~al.}}]{CLEO1999.3746}%
  \BibitemOpen
  \bibfield  {author} {\bibinfo {author} {\bibfnamefont {J.}~\bibnamefont
  {Bartelt}} \emph {et~al.} (\bibinfo {collaboration} {CLEO Collaboration}),\
  }\href {\doibase 10.1103/PhysRevLett.82.3746} {\bibfield  {journal} {\bibinfo
   {journal} {Physical Review Letters}\ }\textbf {\bibinfo {volume} {82}},\
  \bibinfo {pages} {3746} (\bibinfo {year} {1999})}\BibitemShut {NoStop}%
\bibitem [{\citenamefont {Abbiendi}\ \emph {et~al.}(2000)\citenamefont
  {Abbiendi} \emph {et~al.}}]{OPAL2000.15}%
  \BibitemOpen
  \bibfield  {author} {\bibinfo {author} {\bibfnamefont {G.}~\bibnamefont
  {Abbiendi}} \emph {et~al.} (\bibinfo {collaboration} {OPAL Collaboration}),\
  }\href {\doibase https://doi.org/10.1016/S0370-2693(00)00457-3} {\bibfield
  {journal} {\bibinfo  {journal} {Physics Letters B}\ }\textbf {\bibinfo
  {volume} {482}},\ \bibinfo {pages} {15 } (\bibinfo {year}
  {2000})}\BibitemShut {NoStop}%
\bibitem [{\citenamefont {Abe}\ \emph {et~al.}(2002)\citenamefont {Abe} \emph
  {et~al.}}]{Belle2002.258}%
  \BibitemOpen
  \bibfield  {author} {\bibinfo {author} {\bibfnamefont {K.}~\bibnamefont
  {Abe}} \emph {et~al.} (\bibinfo {collaboration} {Belle Collaboration}),\
  }\href {\doibase 10.1016/S0370-2693(01)01483-6} {\bibfield  {journal}
  {\bibinfo  {journal} {Physics Letters B}\ }\textbf {\bibinfo {volume}
  {526}},\ \bibinfo {pages} {258} (\bibinfo {year} {2002})}\BibitemShut
  {NoStop}%
\bibitem [{\citenamefont {Aubert}\ \emph
  {et~al.}(2008{\natexlab{a}})\citenamefont {Aubert} \emph
  {et~al.}}]{BABAR2008.032002}%
  \BibitemOpen
  \bibfield  {author} {\bibinfo {author} {\bibfnamefont {B.}~\bibnamefont
  {Aubert}} \emph {et~al.} (\bibinfo {collaboration} {BABAR Collaboration}),\
  }\href {\doibase 10.1103/PhysRevD.77.032002} {\bibfield  {journal} {\bibinfo
  {journal} {Physical Review D}\ }\textbf {\bibinfo {volume} {77}},\ \bibinfo
  {pages} {032002} (\bibinfo {year} {2008}{\natexlab{a}})}\BibitemShut
  {NoStop}%
\bibitem [{\citenamefont {Aubert}\ \emph {et~al.}(2010)\citenamefont {Aubert}
  \emph {et~al.}}]{BABAR2010.011802}%
  \BibitemOpen
  \bibfield  {author} {\bibinfo {author} {\bibfnamefont {B.}~\bibnamefont
  {Aubert}} \emph {et~al.} (\bibinfo {collaboration} {BABAR Collaboration}),\
  }\href {\doibase 10.1103/PhysRevLett.104.011802} {\bibfield  {journal}
  {\bibinfo  {journal} {Physical Review Letters}\ }\textbf {\bibinfo {volume}
  {104}},\ \bibinfo {pages} {011802} (\bibinfo {year} {2010})}\BibitemShut
  {NoStop}%
\bibitem [{\citenamefont {Dungel}\ \emph {et~al.}(2010)\citenamefont {Dungel},
  \citenamefont {Aziz} \emph {et~al.}}]{Belle2010.112007}%
  \BibitemOpen
  \bibfield  {author} {\bibinfo {author} {\bibfnamefont {W.}~\bibnamefont
  {Dungel}}, \bibinfo {author} {\bibfnamefont {T.}~\bibnamefont {Aziz}},  \emph
  {et~al.} (\bibinfo {collaboration} {Belle Collaboration}),\ }\href {\doibase
  10.1103/PhysRevD.82.112007} {\bibfield  {journal} {\bibinfo  {journal}
  {Physical Review D}\ }\textbf {\bibinfo {volume} {82}},\ \bibinfo {pages}
  {112007} (\bibinfo {year} {2010})}\BibitemShut {NoStop}%
\bibitem [{\citenamefont {Glattauer}\ \emph {et~al.}(2016)\citenamefont
  {Glattauer}, \citenamefont {Schwanda} \emph {et~al.}}]{Belle2016.032006}%
  \BibitemOpen
  \bibfield  {author} {\bibinfo {author} {\bibfnamefont {R.}~\bibnamefont
  {Glattauer}}, \bibinfo {author} {\bibnamefont {Schwanda}},  \emph {et~al.}
  (\bibinfo {collaboration} {Belle Collaboration}),\ }\href {\doibase
  10.1103/PhysRevD.93.032006} {\bibfield  {journal} {\bibinfo  {journal}
  {Physical Review D}\ }\textbf {\bibinfo {volume} {93}},\ \bibinfo {pages}
  {032006} (\bibinfo {year} {2016})}\BibitemShut {NoStop}%
\bibitem [{\citenamefont {Aubert}\ \emph
  {et~al.}(2008{\natexlab{b}})\citenamefont {Aubert} \emph
  {et~al.}}]{BABAR2008}%
  \BibitemOpen
  \bibfield  {author} {\bibinfo {author} {\bibfnamefont {B.}~\bibnamefont
  {Aubert}} \emph {et~al.} (\bibinfo {collaboration} {The BABAR
  Collaboration}),\ }\href {\doibase 10.1103/PhysRevLett.101.261802} {\bibfield
   {journal} {\bibinfo  {journal} {Phys. Rev. Lett.}\ }\textbf {\bibinfo
  {volume} {101}},\ \bibinfo {pages} {261802} (\bibinfo {year}
  {2008}{\natexlab{b}})}\BibitemShut {NoStop}%
\bibitem [{\citenamefont {Liventsev}\ \emph {et~al.}(2008)\citenamefont
  {Liventsev} \emph {et~al.}}]{Belle2008}%
  \BibitemOpen
  \bibfield  {author} {\bibinfo {author} {\bibfnamefont {D.}~\bibnamefont
  {Liventsev}} \emph {et~al.} (\bibinfo {collaboration} {The Belle
  Collaboration}),\ }\href {\doibase 10.1103/PhysRevD.77.091503} {\bibfield
  {journal} {\bibinfo  {journal} {Phys. Rev. D}\ }\textbf {\bibinfo {volume}
  {77}},\ \bibinfo {pages} {091503} (\bibinfo {year} {2008})}\BibitemShut
  {NoStop}%
\bibitem [{\citenamefont {{K. Abe \emph{et al.}}}(2004)}]{Belle2004}%
  \BibitemOpen
  \bibfield  {author} {\bibinfo {author} {\bibnamefont {{K. Abe \emph{et
  al.}}}} (\bibinfo {collaboration} {Belle Collaboration}),\ }\href {\doibase
  10.1103/PhysRevD.69.112002} {\bibfield  {journal} {\bibinfo  {journal} {Phys.
  Rev. D}\ }\textbf {\bibinfo {volume} {69}},\ \bibinfo {pages} {112002}
  (\bibinfo {year} {2004})}\BibitemShut {NoStop}%
\bibitem [{\citenamefont {Kuzmin}\ \emph {et~al.}(2007)\citenamefont {Kuzmin}
  \emph {et~al.}}]{Belle2007.012006}%
  \BibitemOpen
  \bibfield  {author} {\bibinfo {author} {\bibfnamefont {A.}~\bibnamefont
  {Kuzmin}} \emph {et~al.} (\bibinfo {collaboration} {Belle Collaboration}),\
  }\href {\doibase 10.1103/PhysRevD.76.012006} {\bibfield  {journal} {\bibinfo
  {journal} {Phys. Rev. D}\ }\textbf {\bibinfo {volume} {76}},\ \bibinfo
  {pages} {012006} (\bibinfo {year} {2007})}\BibitemShut {NoStop}%
\bibitem [{\citenamefont {Aubert}\ \emph {et~al.}(2009)\citenamefont {Aubert},
  \citenamefont {Bona} \emph {et~al.}}]{BABAR2009.112004}%
  \BibitemOpen
  \bibfield  {author} {\bibinfo {author} {\bibfnamefont {B.}~\bibnamefont
  {Aubert}}, \bibinfo {author} {\bibfnamefont {M.}~\bibnamefont {Bona}},  \emph
  {et~al.} (\bibinfo {collaboration} {BABAR Collaboration}),\ }\href {\doibase
  10.1103/PhysRevD.79.112004} {\bibfield  {journal} {\bibinfo  {journal}
  {Physical Review D}\ }\textbf {\bibinfo {volume} {79}},\ \bibinfo {pages}
  {112004} (\bibinfo {year} {2009})},\ \Eprint {http://arxiv.org/abs/0901.1291}
  {arXiv:0901.1291} \BibitemShut {NoStop}%
\bibitem [{\citenamefont {Aaij}\ \emph
  {et~al.}(2015{\natexlab{a}})\citenamefont {Aaij} \emph
  {et~al.}}]{LHCb2015.032002}%
  \BibitemOpen
  \bibfield  {author} {\bibinfo {author} {\bibfnamefont {R.}~\bibnamefont
  {Aaij}} \emph {et~al.} (\bibinfo {collaboration} {LHCb Collaboration}),\
  }\href {\doibase 10.1103/PhysRevD.92.032002} {\bibfield  {journal} {\bibinfo
  {journal} {Phys. Rev. D}\ }\textbf {\bibinfo {volume} {92}},\ \bibinfo
  {pages} {032002} (\bibinfo {year} {2015}{\natexlab{a}})}\BibitemShut
  {NoStop}%
\bibitem [{\citenamefont {Aaij}\ \emph
  {et~al.}(2015{\natexlab{b}})\citenamefont {Aaij} \emph
  {et~al.}}]{LHCb2015.092002}%
  \BibitemOpen
  \bibfield  {author} {\bibinfo {author} {\bibfnamefont {R.}~\bibnamefont
  {Aaij}} \emph {et~al.} (\bibinfo {collaboration} {LHCb Collaboration}),\
  }\href {\doibase 10.1103/PhysRevD.91.092002} {\bibfield  {journal} {\bibinfo
  {journal} {Phys. Rev. D}\ }\textbf {\bibinfo {volume} {91}},\ \bibinfo
  {pages} {092002} (\bibinfo {year} {2015}{\natexlab{b}})}\BibitemShut
  {NoStop}%
\bibitem [{\citenamefont {Kang}\ \emph {et~al.}(2018)\citenamefont {Kang},
  \citenamefont {Luo}, \citenamefont {Zhang}, \citenamefont {Dai},\ and\
  \citenamefont {Wang}}]{Kang2018}%
  \BibitemOpen
  \bibfield  {author} {\bibinfo {author} {\bibfnamefont {X.-W.}\ \bibnamefont
  {Kang}}, \bibinfo {author} {\bibfnamefont {T.}~\bibnamefont {Luo}}, \bibinfo
  {author} {\bibfnamefont {Y.}~\bibnamefont {Zhang}}, \bibinfo {author}
  {\bibfnamefont {L.-Y.}\ \bibnamefont {Dai}}, \ and\ \bibinfo {author}
  {\bibfnamefont {C.}~\bibnamefont {Wang}},\ }\href {\doibase
  10.1140/epjc/s10052-018-6385-9} {\bibfield  {journal} {\bibinfo  {journal}
  {The European Physical Journal C}\ }\textbf {\bibinfo {volume} {78}},\
  \bibinfo {pages} {909} (\bibinfo {year} {2018})}\BibitemShut {NoStop}%
\bibitem [{\citenamefont {Segovia}\ \emph {et~al.}(2011)\citenamefont
  {Segovia}, \citenamefont {Albertus}, \citenamefont {Entem}, \citenamefont
  {Fern{\'{a}}ndez}, \citenamefont {Hern{\'{a}}ndez},\ and\ \citenamefont
  {P{\'{e}}rez-Garc{\'{i}}a}}]{Segovia2011.094029}%
  \BibitemOpen
  \bibfield  {author} {\bibinfo {author} {\bibfnamefont {J.}~\bibnamefont
  {Segovia}}, \bibinfo {author} {\bibfnamefont {C.}~\bibnamefont {Albertus}},
  \bibinfo {author} {\bibfnamefont {D.~R.}\ \bibnamefont {Entem}}, \bibinfo
  {author} {\bibfnamefont {F.}~\bibnamefont {Fern{\'{a}}ndez}}, \bibinfo
  {author} {\bibfnamefont {E.}~\bibnamefont {Hern{\'{a}}ndez}}, \ and\ \bibinfo
  {author} {\bibfnamefont {M.~A.}\ \bibnamefont {P{\'{e}}rez-Garc{\'{i}}a}},\
  }\href {\doibase 10.1103/PhysRevD.84.094029} {\bibfield  {journal} {\bibinfo
  {journal} {Physical Review D}\ }\textbf {\bibinfo {volume} {84}},\ \bibinfo
  {pages} {094029} (\bibinfo {year} {2011})}\BibitemShut {NoStop}%
\bibitem [{\citenamefont {Fu}\ \emph {et~al.}(2011)\citenamefont {Fu},
  \citenamefont {Wang}, \citenamefont {Wang},\ and\ \citenamefont
  {Chen}}]{FuHF2011}%
  \BibitemOpen
  \bibfield  {author} {\bibinfo {author} {\bibfnamefont {H.-F.}\ \bibnamefont
  {Fu}}, \bibinfo {author} {\bibfnamefont {G.-L.}\ \bibnamefont {Wang}},
  \bibinfo {author} {\bibfnamefont {Z.-H.}\ \bibnamefont {Wang}}, \ and\
  \bibinfo {author} {\bibfnamefont {X.-J.}\ \bibnamefont {Chen}},\ }\href
  {\doibase 10.1088/0256-307X/28/12/121301} {\bibfield  {journal} {\bibinfo
  {journal} {Chinese Physics Letters}\ }\textbf {\bibinfo {volume} {28}},\
  \bibinfo {pages} {121301} (\bibinfo {year} {2011})}\BibitemShut {NoStop}%
\bibitem [{\citenamefont {{R. Aaij \emph{et al.}}}(2013)}]{LHCb2013}%
  \BibitemOpen
  \bibfield  {author} {\bibinfo {author} {\bibnamefont {{R. Aaij \emph{et
  al.}}}} (\bibinfo {collaboration} {LHCb Collaboration}),\ }\href {\doibase
  10.1007/JHEP09(2013)145} {\bibfield  {journal} {\bibinfo  {journal} {JHEP}\
  }\textbf {\bibinfo {volume} {2013}},\ \bibinfo {pages} {145} (\bibinfo {year}
  {2013})}\BibitemShut {NoStop}%
\bibitem [{\citenamefont {Sun}\ \emph {et~al.}(2013)\citenamefont {Sun},
  \citenamefont {Liu},\ and\ \citenamefont {Matsuki}}]{Sun2013}%
  \BibitemOpen
  \bibfield  {author} {\bibinfo {author} {\bibfnamefont {Y.}~\bibnamefont
  {Sun}}, \bibinfo {author} {\bibfnamefont {X.}~\bibnamefont {Liu}}, \ and\
  \bibinfo {author} {\bibfnamefont {T.}~\bibnamefont {Matsuki}},\ }\href
  {\doibase 10.1103/PhysRevD.88.094020} {\bibfield  {journal} {\bibinfo
  {journal} {Phys. Rev. D}\ }\textbf {\bibinfo {volume} {88}},\ \bibinfo
  {pages} {094020} (\bibinfo {year} {2013})}\BibitemShut {NoStop}%
\bibitem [{\citenamefont {L{\"{u}}}\ and\ \citenamefont {Li}(2014)}]{Lu2014}%
  \BibitemOpen
  \bibfield  {author} {\bibinfo {author} {\bibfnamefont {Q.-F.}\ \bibnamefont
  {L{\"{u}}}}\ and\ \bibinfo {author} {\bibfnamefont {D.-M.}\ \bibnamefont
  {Li}},\ }\href {\doibase 10.1103/PhysRevD.90.054024} {\bibfield  {journal}
  {\bibinfo  {journal} {Phys. Rev. D}\ }\textbf {\bibinfo {volume} {90}},\
  \bibinfo {pages} {054024} (\bibinfo {year} {2014})}\BibitemShut {NoStop}%
\bibitem [{\citenamefont {Yu}\ \emph {et~al.}(2015)\citenamefont {Yu},
  \citenamefont {Wang}, \citenamefont {Li},\ and\ \citenamefont
  {Meng}}]{YuGL2015}%
  \BibitemOpen
  \bibfield  {author} {\bibinfo {author} {\bibfnamefont {G.-L.}\ \bibnamefont
  {Yu}}, \bibinfo {author} {\bibfnamefont {Z.-G.}\ \bibnamefont {Wang}},
  \bibinfo {author} {\bibfnamefont {Z.-Y.}\ \bibnamefont {Li}}, \ and\ \bibinfo
  {author} {\bibfnamefont {G.-Q.}\ \bibnamefont {Meng}},\ }\href
  {http://stacks.iop.org/1674-1137/39/i=6/a=063101} {\bibfield  {journal}
  {\bibinfo  {journal} {Chin. Phys. C}\ }\textbf {\bibinfo {volume} {39}},\
  \bibinfo {pages} {063101} (\bibinfo {year} {2015})}\BibitemShut {NoStop}%
\bibitem [{\citenamefont {Godfrey}\ and\ \citenamefont
  {Moats}(2016)}]{Godfrey2016}%
  \BibitemOpen
  \bibfield  {author} {\bibinfo {author} {\bibfnamefont {S.}~\bibnamefont
  {Godfrey}}\ and\ \bibinfo {author} {\bibfnamefont {K.}~\bibnamefont
  {Moats}},\ }\href {\doibase 10.1103/PhysRevD.93.034035} {\bibfield  {journal}
  {\bibinfo  {journal} {Phys. Rev.D}\ }\textbf {\bibinfo {volume} {93}},\
  \bibinfo {pages} {034035} (\bibinfo {year} {2016})}\BibitemShut {NoStop}%
\bibitem [{\citenamefont {Li}\ \emph {et~al.}(2017{\natexlab{a}})\citenamefont
  {Li}, \citenamefont {Jiang}, \citenamefont {Wang}, \citenamefont {Li},
  \citenamefont {Wang},\ and\ \citenamefont {Wang}}]{lsc2017}%
  \BibitemOpen
  \bibfield  {author} {\bibinfo {author} {\bibfnamefont {S.-C.}\ \bibnamefont
  {Li}}, \bibinfo {author} {\bibfnamefont {Y.}~\bibnamefont {Jiang}}, \bibinfo
  {author} {\bibfnamefont {T.-H.}\ \bibnamefont {Wang}}, \bibinfo {author}
  {\bibfnamefont {Q.}~\bibnamefont {Li}}, \bibinfo {author} {\bibfnamefont
  {Z.-H.}\ \bibnamefont {Wang}}, \ and\ \bibinfo {author} {\bibfnamefont
  {G.-L.}\ \bibnamefont {Wang}},\ }\href {\doibase 10.1142/S0217732317500134}
  {\bibfield  {journal} {\bibinfo  {journal} {Mod. Phys. Lett. A}\ }\textbf
  {\bibinfo {volume} {32}},\ \bibinfo {pages} {1750013} (\bibinfo {year}
  {2017}{\natexlab{a}})}\BibitemShut {NoStop}%
\bibitem [{\citenamefont {Li}\ \emph {et~al.}(2018)\citenamefont {Li},
  \citenamefont {Wang}, \citenamefont {Jiang}, \citenamefont {Tan},
  \citenamefont {Li}, \citenamefont {Wang},\ and\ \citenamefont
  {Chang}}]{lsc2018}%
  \BibitemOpen
  \bibfield  {author} {\bibinfo {author} {\bibfnamefont {S.-C.}\ \bibnamefont
  {Li}}, \bibinfo {author} {\bibfnamefont {T.}~\bibnamefont {Wang}}, \bibinfo
  {author} {\bibfnamefont {Y.}~\bibnamefont {Jiang}}, \bibinfo {author}
  {\bibfnamefont {X.-Z.}\ \bibnamefont {Tan}}, \bibinfo {author} {\bibfnamefont
  {Q.}~\bibnamefont {Li}}, \bibinfo {author} {\bibfnamefont {G.-L.}\
  \bibnamefont {Wang}}, \ and\ \bibinfo {author} {\bibfnamefont {C.-H.}\
  \bibnamefont {Chang}},\ }\href {\doibase 10.1103/PhysRevD.97.054002}
  {\bibfield  {journal} {\bibinfo  {journal} {Phys. Rev. D}\ }\textbf {\bibinfo
  {volume} {97}},\ \bibinfo {pages} {054002} (\bibinfo {year}
  {2018})}\BibitemShut {NoStop}%
\bibitem [{\citenamefont {Tan}\ \emph {et~al.}(2018)\citenamefont {Tan},
  \citenamefont {Wang}, \citenamefont {Jiang}, \citenamefont {Li},
  \citenamefont {Li}, \citenamefont {Wang},\ and\ \citenamefont
  {Chang}}]{Tan2018}%
  \BibitemOpen
  \bibfield  {author} {\bibinfo {author} {\bibfnamefont {X.-Z.}\ \bibnamefont
  {Tan}}, \bibinfo {author} {\bibfnamefont {T.}~\bibnamefont {Wang}}, \bibinfo
  {author} {\bibfnamefont {Y.}~\bibnamefont {Jiang}}, \bibinfo {author}
  {\bibfnamefont {S.-C.}\ \bibnamefont {Li}}, \bibinfo {author} {\bibfnamefont
  {Q.}~\bibnamefont {Li}}, \bibinfo {author} {\bibfnamefont {G.-L.}\
  \bibnamefont {Wang}}, \ and\ \bibinfo {author} {\bibfnamefont {C.-H.}\
  \bibnamefont {Chang}},\ }\href {\doibase 10.1140/epjc/s10052-018-6054-z}
  {\bibfield  {journal} {\bibinfo  {journal} {The European Physical Journal C}\
  }\textbf {\bibinfo {volume} {78}},\ \bibinfo {pages} {583} (\bibinfo {year}
  {2018})}\BibitemShut {NoStop}%
\bibitem [{\citenamefont {Geng}\ \emph {et~al.}(2019)\citenamefont {Geng},
  \citenamefont {Wang}, \citenamefont {Jiang}, \citenamefont {Li},
  \citenamefont {Tan},\ and\ \citenamefont {Wang}}]{gzk2019}%
  \BibitemOpen
  \bibfield  {author} {\bibinfo {author} {\bibfnamefont {Z.-K.}\ \bibnamefont
  {Geng}}, \bibinfo {author} {\bibfnamefont {T.}~\bibnamefont {Wang}}, \bibinfo
  {author} {\bibfnamefont {Y.}~\bibnamefont {Jiang}}, \bibinfo {author}
  {\bibfnamefont {G.}~\bibnamefont {Li}}, \bibinfo {author} {\bibfnamefont
  {X.-Z.}\ \bibnamefont {Tan}}, \ and\ \bibinfo {author} {\bibfnamefont
  {G.-L.}\ \bibnamefont {Wang}},\ }\href {\doibase 10.1103/PhysRevD.99.013006}
  {\bibfield  {journal} {\bibinfo  {journal} {Phys. Rev. D}\ }\textbf {\bibinfo
  {volume} {99}},\ \bibinfo {pages} {013006} (\bibinfo {year}
  {2019})}\BibitemShut {NoStop}%
\bibitem [{\citenamefont {Kim}\ and\ \citenamefont {Wang}(2004)}]{WangGL2004}%
  \BibitemOpen
  \bibfield  {author} {\bibinfo {author} {\bibfnamefont {C.~S.}\ \bibnamefont
  {Kim}}\ and\ \bibinfo {author} {\bibfnamefont {G.-L.}\ \bibnamefont {Wang}},\
  }\href {\doibase 10.1016/j.physletb.2004.01.058} {\bibfield  {journal}
  {\bibinfo  {journal} {Phys. Lett.}\ }\textbf {\bibinfo {volume} {B584}},\
  \bibinfo {pages} {285} (\bibinfo {year} {2004})}\BibitemShut {NoStop}%
\bibitem [{\citenamefont {Wang}(2006)}]{WangGL2006}%
  \BibitemOpen
  \bibfield  {author} {\bibinfo {author} {\bibfnamefont {G.-L.}\ \bibnamefont
  {Wang}},\ }\href {\doibase 10.1016/j.physletb.2005.12.005} {\bibfield
  {journal} {\bibinfo  {journal} {Phys. Lett. B}\ }\textbf {\bibinfo {volume}
  {633}},\ \bibinfo {pages} {492} (\bibinfo {year} {2006})}\BibitemShut
  {NoStop}%
\bibitem [{\citenamefont {Wang}(2007)}]{WangGL2007}%
  \BibitemOpen
  \bibfield  {author} {\bibinfo {author} {\bibfnamefont {G.-L.}\ \bibnamefont
  {Wang}},\ }\href {\doibase 10.1016/j.physletb.2007.05.001} {\bibfield
  {journal} {\bibinfo  {journal} {Physics Letters B}\ }\textbf {\bibinfo
  {volume} {650}},\ \bibinfo {pages} {15} (\bibinfo {year} {2007})}\BibitemShut
  {NoStop}%
\bibitem [{\citenamefont {Ali}\ \emph {et~al.}(1998)\citenamefont {Ali},
  \citenamefont {Kramer},\ and\ \citenamefont {L\"u}}]{Ali1998.094009}%
  \BibitemOpen
  \bibfield  {author} {\bibinfo {author} {\bibfnamefont {A.}~\bibnamefont
  {Ali}}, \bibinfo {author} {\bibfnamefont {G.}~\bibnamefont {Kramer}}, \ and\
  \bibinfo {author} {\bibfnamefont {C.-D.}\ \bibnamefont {L\"u}},\ }\href
  {\doibase 10.1103/PhysRevD.58.094009} {\bibfield  {journal} {\bibinfo
  {journal} {Phys. Rev. D}\ }\textbf {\bibinfo {volume} {58}},\ \bibinfo
  {pages} {094009} (\bibinfo {year} {1998})}\BibitemShut {NoStop}%
\bibitem [{\citenamefont {Choi}\ and\ \citenamefont
  {Ji}(2009)}]{Choi2009.114003}%
  \BibitemOpen
  \bibfield  {author} {\bibinfo {author} {\bibfnamefont {H.-M.}\ \bibnamefont
  {Choi}}\ and\ \bibinfo {author} {\bibfnamefont {C.-R.}\ \bibnamefont {Ji}},\
  }\href {\doibase 10.1103/PhysRevD.80.114003} {\bibfield  {journal} {\bibinfo
  {journal} {Phys. Rev. D}\ }\textbf {\bibinfo {volume} {80}},\ \bibinfo
  {pages} {114003} (\bibinfo {year} {2009})}\BibitemShut {NoStop}%
\bibitem [{\citenamefont {Fakirov}\ and\ \citenamefont
  {Stech}(1978)}]{Fakirov1978.315}%
  \BibitemOpen
  \bibfield  {author} {\bibinfo {author} {\bibfnamefont {D.}~\bibnamefont
  {Fakirov}}\ and\ \bibinfo {author} {\bibfnamefont {B.}~\bibnamefont
  {Stech}},\ }\href {\doibase 10.1016/0550-3213(78)90306-1} {\bibfield
  {journal} {\bibinfo  {journal} {Nuclear Physics B}\ }\textbf {\bibinfo
  {volume} {133}},\ \bibinfo {pages} {315 } (\bibinfo {year}
  {1978})}\BibitemShut {NoStop}%
\bibitem [{\citenamefont {Bauer}\ \emph {et~al.}(1987)\citenamefont {Bauer},
  \citenamefont {Stech},\ and\ \citenamefont {Wirbel}}]{Bauer1987.103}%
  \BibitemOpen
  \bibfield  {author} {\bibinfo {author} {\bibfnamefont {M.}~\bibnamefont
  {Bauer}}, \bibinfo {author} {\bibfnamefont {B.}~\bibnamefont {Stech}}, \ and\
  \bibinfo {author} {\bibfnamefont {M.}~\bibnamefont {Wirbel}},\ }\href
  {\doibase 10.1007/BF01561122} {\bibfield  {journal} {\bibinfo  {journal}
  {Zeitschrift f{\"u}r Physik C Particles and Fields}\ }\textbf {\bibinfo
  {volume} {34}},\ \bibinfo {pages} {103} (\bibinfo {year} {1987})}\BibitemShut
  {NoStop}%
\bibitem [{\citenamefont {Ivanov}\ \emph {et~al.}(2006)\citenamefont {Ivanov},
  \citenamefont {K{\"{o}}rner},\ and\ \citenamefont
  {Santorelli}}]{Ivanov2006.054024}%
  \BibitemOpen
  \bibfield  {author} {\bibinfo {author} {\bibfnamefont {M.~A.}\ \bibnamefont
  {Ivanov}}, \bibinfo {author} {\bibfnamefont {J.~G.}\ \bibnamefont
  {K{\"{o}}rner}}, \ and\ \bibinfo {author} {\bibfnamefont {P.}~\bibnamefont
  {Santorelli}},\ }\href {\doibase 10.1103/PhysRevD.73.054024} {\bibfield
  {journal} {\bibinfo  {journal} {Physical Review D}\ }\textbf {\bibinfo
  {volume} {73}},\ \bibinfo {pages} {054024} (\bibinfo {year}
  {2006})}\BibitemShut {NoStop}%
\bibitem [{\citenamefont {Tanabashi}\ \emph {et~al.}(2018)\citenamefont
  {Tanabashi} \emph {et~al.}}]{PDG2018}%
  \BibitemOpen
  \bibfield  {author} {\bibinfo {author} {\bibfnamefont {M.}~\bibnamefont
  {Tanabashi}} \emph {et~al.} (\bibinfo {collaboration} {Particle Data
  Group}),\ }\href {\doibase 10.1103/PhysRevD.98.030001} {\bibfield  {journal}
  {\bibinfo  {journal} {Phys. Rev. D}\ }\textbf {\bibinfo {volume} {98}},\
  \bibinfo {pages} {030001} (\bibinfo {year} {2018})}\BibitemShut {NoStop}%
\bibitem [{\citenamefont {Li}\ \emph {et~al.}(2017{\natexlab{b}})\citenamefont
  {Li}, \citenamefont {Jiang}, \citenamefont {Wang}, \citenamefont {Yuan},
  \citenamefont {Wang},\ and\ \citenamefont {Chang}}]{LiQ2017}%
  \BibitemOpen
  \bibfield  {author} {\bibinfo {author} {\bibfnamefont {Q.}~\bibnamefont
  {Li}}, \bibinfo {author} {\bibfnamefont {Y.}~\bibnamefont {Jiang}}, \bibinfo
  {author} {\bibfnamefont {T.}~\bibnamefont {Wang}}, \bibinfo {author}
  {\bibfnamefont {H.}~\bibnamefont {Yuan}}, \bibinfo {author} {\bibfnamefont
  {G.-L.}\ \bibnamefont {Wang}}, \ and\ \bibinfo {author} {\bibfnamefont
  {C.-H.}\ \bibnamefont {Chang}},\ }\href {\doibase
  10.1140/epjc/s10052-017-4865-y} {\bibfield  {journal} {\bibinfo  {journal}
  {Eur. Phys. J. C}\ }\textbf {\bibinfo {volume} {77}},\ \bibinfo {pages} {297}
  (\bibinfo {year} {2017}{\natexlab{b}})}\BibitemShut {NoStop}%
\bibitem [{\citenamefont {Aaij}\ \emph {et~al.}(2016)\citenamefont {Aaij} \emph
  {et~al.}}]{LHCb2016.119901}%
  \BibitemOpen
  \bibfield  {author} {\bibinfo {author} {\bibfnamefont {R.}~\bibnamefont
  {Aaij}} \emph {et~al.} (\bibinfo {collaboration} {LHCb Collaboration}),\
  }\href {\doibase 10.1103/PhysRevD.93.119901} {\bibfield  {journal} {\bibinfo
  {journal} {Phys. Rev. D}\ }\textbf {\bibinfo {volume} {93}},\ \bibinfo
  {pages} {119901} (\bibinfo {year} {2016})}\BibitemShut {NoStop}%
\end{thebibliography}%

%%or paste the xxx.bbl file(generated from bibtex) here .
%%%%%%%%%%%%%%%%%%%%%%%%%%%%%%%%%%%%%%%%%%%%%%%%%%%%%%%%%%%%

%%%%%%%%%%%%%%%%%%%%%%%%%%%%%%%%%%%%%%%%%%%%%%%%%%%%%%%%%%%%%%%%%%%%%%%%%

\end{document}